\documentclass[useAMS,fleqn,usenatbib]{mnras}

\usepackage[T1]{fontenc}
\usepackage{ae,aecompl}

\usepackage{amsmath}
\usepackage{amsopn}
\usepackage[british]{babel}
\usepackage[varg]{txfonts}
\usepackage{biblio} 
\usepackage{natbib}
\usepackage{color}

\usepackage{graphicx}
\usepackage{epstopdf} 

\usepackage{microtype}

\definecolor{orange}{rgb}{1.0,0.5,0.}

\DeclareMathOperator{\sech}{sech}

\def\MDM{\ifmmode{\>M_{\textnormal{\sc dm}}}\else{$$M_{\textnormal{\sc dm}}}\fi}

\def\XH{\ifmmode{\>X_{\textnormal{\sc h}}} \else{$X_{\textnormal{\sc h}}$}\fi}
\def\nH{\ifmmode{\>n_{\textnormal{\sc h}}} \else{$n_{\textnormal{\sc h}}$}\fi}

\def\maspyr{\ifmmode{\>\textnormal{mas~yr}^{-1}}\else{mas~yr$^{-1}$}\fi}

\def\mG{\ifmmode{\>\mu\mathrm{G}}\else{$\mu$G}\fi}
\def\erg{\ifmmode{\> {\rm erg}}\else{erg}\fi}
\def\keV{\ifmmode{\> {\rm keV}}\else{keV}\fi}

\def\deg{\ifmmode{\>^{\circ}}\else{$^{\circ}$}\fi}
\def\onedeg{\ifmmode{\>1^{\circ}}\else{$1^{\circ}$}\fi}

\def\xvir{\ifmmode{\>\!x_{vir}}\else{$x_{vir}$}\fi}
\def\Mvir{\ifmmode{\>\!M_{vir} }\else{$M_{vir} $}\fi}
\def\rvir{\ifmmode{\>\!r_{vir}}\else{$r_{vir}$}\fi}
\def\vvir{\ifmmode{\>\!v_{vir}}\else{$v_{vir}$}\fi}
\def\Vvir{\ifmmode{\>\!V_{vir} }\else{$V_{vir} $}\fi}

\def\tratio{\ifmmode{\>\tau}\else{$\tau$}\fi}

\def\rms{\ifmmode{\>r_{\textnormal{\sc ms}}}\else{$r_{\textnormal{\sc ms}}$}\fi}

\def\Mpc{\ifmmode{\>\!{\rm Mpc}} \else{Mpc}\fi}
\def\kpc{\ifmmode{\>\!{\rm kpc}} \else{kpc}\fi}
\def\pkpc{\ifmmode{\>\!{\rm kpc}^{-1}} \else{kpc$^{-1}$}\fi}
\def\pc{\ifmmode{\>\!{\rm pc}} \else{pc}\fi}

\def\Gyr{\ifmmode{\>\!{\rm Gyr}} \else{Gyr}\fi}
\def\Myr{\ifmmode{\>\!{\rm Myr}} \else{Myr}\fi}
\def\yr{\ifmmode{\>\!{\rm yr}} \else{yr}\fi}
\def\pyr{\ifmmode{\>\!{\rm yr}^{-1}}\else{yr$^{-1}$} \fi}
\def\s{\ifmmode{\>\!{\rm s}}\else{s}\fi}
\def\ps{\ifmmode{\>\!{\rm s}^{-1}}\else{s$^{-1}$}\fi}
\def\Hz{\ifmmode{\>\!{\rm Hz}}\else{Hz}\fi}

\def\kms{\ifmmode{\>\!{\rm km\,s}^{-1}}\else{km~s$^{-1}$}\fi}

\def\K{\ifmmode{\>\!{\rm K}}\else{K}\fi}

\def\sr{\ifmmode{\>\!{\rm sr}}\else{sr}\fi}
\def\psr{\ifmmode{\>\!{\rm sr}^{-1}}\else{sr$^{-1}$}\fi}
\def\arcs{\ifmmode{\>\!{\rm arcsec}}\else{arcsec}\fi}
\def\parcs{\ifmmode{\>\!{\rm arcsec}^{-1}}\else{arcsec${-1}$}\fi}
\def\parcss{\ifmmode{\>\!{\rm arcsec}^{-2}}\else{arcsec${-2}$}\fi}

\def\cm{\ifmmode{\>\!{\rm cm}}\else{cm}\fi}
\def\cc{\ifmmode{\>\!{\rm cm}^{3}}\else{cm$^{3}$}\fi}
\def\sqc{\ifmmode{\>\!{\rm cm}^{2}}\else{cm$^{2}$}\fi}
\def\pcc{\ifmmode{\>\!{\rm cm}^{-3}}\else{cm$^{-3}$}\fi}
\def\psc{\ifmmode{\>\!{\rm cm}^{-2}}\else{cm$^{-2}$}\fi}

\def\g{\ifmmode{\>\!{\rm g}}\else{g}\fi}
\def\Msun{\ifmmode{\>\!{\rm M}_{\odot}}\else{M$_{\odot}$}\fi}
\def\hMsun{\ifmmode{\> h^{-1}{\rm M}_{\odot}}\else{$h^{-1}$M$_{\odot}$}\fi}

\def\Zsun{\ifmmode{\>\!{\rm Z}_{\odot}}\else{Z$_{\odot}$}\fi}

\def\Lsun{\ifmmode{\>\!{\rm L}_{\odot}}\else{L$_{\odot}$}\fi}

\def\rayl{\ifmmode{\>\!{\rm R}}\else{R}\fi}
\def\mR{\ifmmode{\>\!{\rm mR}}\else{mR}\fi}

\renewcommand{\ion}[2]{\hbox{#1\,{\sc #2}}}

\def\lya{\ifmmode{\>\!{\rm Ly}\alpha}\else{Ly$\alpha$}\fi}

\def\Ha{\ifmmode{\>\!{\rm H}\alpha}\else{H$\alpha$}\fi}
\def\Hb{\ifmmode{\>\!{\rm H}\beta}\else{H$\beta$}\fi}

\def\HI{\ifmmode{\> \textnormal{\ion{H}{i}}} \else{\ion{H}{i}}\fi}
\def\HII{\ifmmode{\> \textnormal{\ion{H}{ii}}} \else{\ion{H}{ii}}\fi}
\def\CIV{\ifmmode{\> \textnormal{\ion{C}{iv}}} \else{\ion{C}{iv}}\fi}
\def\SiIV{\ifmmode{\> \textnormal{\ion{S}{iv}}} \else{\ion{Si}{iv}}\fi}

\def\NH{\ifmmode{\> {\rm N}_{\rm H}} \else{N$_{\rm H}$}\fi}
\def\Ng{\ifmmode{\> {\rm N}_{\rm gas}} \else{N$_{\rm gas}$}\fi}
\def\NHI{\ifmmode{\> {\rm N}_{\HI}} \else{N$_{\HI}$}\fi}
\def\MHI{\ifmmode{\> {\rm M}_{ \HI}} \else{M$_{\HI}$}\fi}

\def\mua{\ifmmode{\>\mu_{ \textnormal{\Ha}}}\else{$\mu_{ \textnormal{\Ha}}$}\fi}
\def\alphabha{\ifmmode{\>\alpha_{B}^{(\textnormal{\Ha})}}\else{$\alpha_{B}^{(\textnormal{\Ha})}$}\fi}

\newcommand{\myemail}{tepper@physics.usyd.edu.au}
\newcommand{\ramses}{{\sc ramses}}
\newcommand{\agama}{{\sc agama}}

\defcitealias{bla21e}{BT21}

\title[ Corrugations in stars and gas ]{ Galactic seismology: joint evolution of impact-triggered stellar and gaseous disc corrugations }

\author[Tepper-Garc\'\i a, Bland-Hawthorn
\& Freeman]{%
Thor Tepper-Garc\'\i a,$^{1,2}$\thanks{\myemail} 
Joss Bland-Hawthorn$^{1,2}$ 
and Ken Freeman$^{4}$
\\
$^1$Sydney Institute for Astronomy, School of Physics, University of Sydney, NSW 2006, Australia\\
$^2$Centre of Excellence for All Sky Astrophysics in Three Dimensions (ASTRO-3D), Australia\\
$^4$Mount Stromlo Observatory, Private Bag, Woden, ACT 2611, Australia
}

\date{Accepted ---. Received ---; in original form ---}

\pubyear{\date{year}}

\begin{document}
\label{firstpage}
\pagerange{\pageref{firstpage}--\pageref{lastpage}}
\maketitle

\pdfminorversion=5

\begin{abstract}
Evidence for wave-like corrugations are well established in the Milky Way and in nearby disc galaxies. These were originally detected as a displacement of the interstellar medium about the midplane, either in terms of vertical distance or vertical velocity. Over the past decade, similar patterns have emerged in the Milky Way's stellar disc. We investigate how these vertical waves are triggered by a passing satellite. Using high-resolution N-body/hydrodynamical simulations, we systematically study how the corrugations set up and evolve jointly in the stellar and gaseous discs. We find that the gas corrugations follow the stellar corrugations, i.e. they are initially in phase although, after a few rotation periods (500-700 Myr), the distinct waves separate and thereafter evolve in different ways.
The spatial and kinematic amplitudes (and thus the energy) of the corrugations dampen with time, with the gaseous corrugation settling at a faster rate ($\sim 800$ Myr vs. $\sim 1$ Gyr). In contrast, the vertical energy of individual disc stars is fairly constant throughout the galaxy's evolution. This difference arises because corrugations are an emergent phenomenon supported by the collective, ordered motions of co-spatial ensembles of stars.
We show that the damping of the stellar corrugations can be understood as a consequence of incomplete phase mixing, while the damping of the gaseous corrugations is a natural consequence of the dissipative nature of the gas.
We suggest that  -- in the absence of further, strong perturbations -- the degree of correlation between the stellar and gaseous waves may help to age-date the phenomenon.

\end{abstract}

\begin{keywords}
methods: analytic -- Surveys -- the Galaxy -- stars: kinematics and dynamics -- methods: N-body simulations -- methods: hydrodynamical simulations
\end{keywords}

\section{Introduction} \label{s:intro}

Galactic discs are a remarkably complex phenomenon that exhibit many distinct properties. These include spiral arms, central bars, bimodal populations, resonant structures (e.g. rings) and outer warps \citep{gil83a,bri90t,but96o},
all of which continue to be important research topics in modern astrophysics \citep{bin08a}.
One property of late-type galaxies that has long been recognised, but gets much less attention, are disc corrugations, essentially a ripple pattern for which the wave amplitude is vertical to the galactic plane.

The earliest evidence of these ripples dates back to the first observations of the outer Galactic warp in HI \citep{bur57r,ker57a,gum60d},
although later observations observed the same effect in molecular gas \citep{san84w}.
These authors noticed a residual wavy structure extending through the disc with respect to the mean Galactic plane.
\citet{qui77s} determined that the ripple pattern increases in wavelength with increasing Galactic radius, as it does in amplitude \citep[][]{gom13l}.
Contemporary observers also noted that young stellar complexes and star-forming regions embedded in the gas appear to show the same wavy behaviour \citep{dix67z,lyn70o,var70n,qui74i,loc77n,alf91q,alf96d}.

Corrugations have been observed in nearby disc galaxies \citep{flo91q,alf01e,sch01s,mat08k,san15h,nar20a}.
So far, there has been one systematic survey based on \Ha\ observations which suggest that 20 percent of nearby spiral galaxies have strongly vertically perturbed discs \citep[][]{urr22t}, but the true relative frequency remains unknown.
Some studies focus on edge-on galaxies where the undulating pattern is particularly evident in the midplane dust lane \citep[e.g.][]{nar20a}. Notably, others detect the corrugation using the kinematic signature associated with the vertical motion \citep[e.g.][]{alf01e,san15h}, which is particularly evident in low-inclination  galaxies \citep[e.g.][]{gom21a}. To date, these studies have identified the undulating patterns almost exclusively through tracers of the multi-phase interstellar medium, including the use of hot young stars associated with HII regions.

To our knowledge, the first detection of corrugations in the {\it stellar} disc arose from Milky Way studies \citep{wid12a,yan13h,xu15v}, 
initially through comparing star counts between the Galactic hemispheres.
\citet{sch18d} provided further evidence using disc kinematics from the Gaia-TGAS data release \citep{gai16r}.\footnote{ \citet{sch18d} make the distinction between propagating wave patterns and breathing modes associated with spiral arms \citep{wil13e,sie12v}. }
More evidence followed on from the ESA Gaia Second Data Release \citep[DR2;][]{gai18m}, specifically the Radial Velocity Survey (RVS) providing 6D phase space information ($x,y,z,V_x,V_y,V_z$) for 6 million stars \citep[e.g.][]{kaw18h,fri19x,xu20x,che19h,lop20v,pog21f,pan21c}, as well as from data collected with the Large Aperture Multi-Object Fiber Spectroscopic Telescope \citep[LAMOST; e.g.][]{car13r,wan20y}. 

So what excites these vertical waves? Theoretical work specific to corrugations dates back to early HI observations \citep{nel76r,nel80v}, although the physics of vertical wave perturbations has a long history \citep{lyn65w,hun69l}.
It was recognised that the outer $m=1$ warp is expected to wrap up with the differential rotation of the disc, and that the gas and stars may evolve differently.

Among the plausible excitation mechanisms are the interaction of the disc with the substructure (subhaloes) within the host halo \citep[e.g.][]{che18a}, misaligned gas accretion \citep{kha22a}, or internal processes \citep[cf.][]{mas96r,kho20g}.
Most of the subsequent literature, however, has focussed on a massive interloper interacting with the disc, although \citet{wei95g} and \citet{elm95t} were the first to recognise the role of orbiting Magellanic-mass galaxies in triggering corrugations \citep[see also][]{ede97b}. These authors observed that the outer warp wraps up into a wavy pattern due to the underlying precession induced by the Galactic potential \citep[cf.][]{bin98f}. We note that, while simulations of disc interactions are commonplace in the literature, the majority of the earlier studies focuses on the evolution of the spiral density wave, the central bar, or the outer disc warp; most do not discuss (or possibly even detect) the corrugation over the inner disc \citep[e.g.][]{kaz08k,you08m,de-15h}.

Yet, in some state-of-the-art, large galaxy simulations, the wave-like pattern is clearly detected and discussed in the context of the outer warp wrapping up with the disc precession, both in controlled, galaxy-scale experiments \citep[e.g.][]{cha09y,gom13l,don16b,lap18a,lap18b,pog21f}, or in full cosmological simulations \citep[e.g.][]{gom16r}. Most recently, \citet[][hereafter \citetalias{bla21e}]{bla21e} simulate the Galactic {\it stellar} disc being disturbed by a single impulse. They separate the evolving spiral density wave from the vertical bending mode, and show that these precess with the disc and wrap up at different rates. They find that the spiral arms `ride the wave' with a distinct ripple pattern along their length. Interestingly, the early HII region studies note that the spiral arms appear to be doing precisely this \citep[e.g.][]{qui77s,kol79q,spi86m}.

In the context of these wave-like patterns, we draw attention to one of the great discoveries of the ESA Gaia satellite, i.e. the remarkable spiral pattern seen in phase space (variously known as the ``snail shell'' or ``phase (space) spiral'') in a local sample of disc stars \citep{ant18b}. The signal is most evident in the $z-V_z$ plane where $z$ is the star’s vertical displacement and $V_z$ is its vertical disc motion. The distribution in this plane can be encoded with another physical property like stellar surface density \citep{lap19a}, angular momentum $L_z$ about the disc rotation axis \citep{kha19a}, or using the other kinematic components \citep{ant18b}. The phase-spiral pattern is seen in all cases.

Gaia's unexpected signal in phase space is further evidence of propagating wave-like patterns in the Galactic disc.
A fast-expanding literature reveals that the phase-spiral pattern can be readily explained by a massive disc-transitting satellite \citep[e.g.][see also \citealt{tia18a,gar22c}]{ant18b,bin18a,lap19a,bla19a,hun21t}.

Disc undulations raise many questions: (i) Are gaseous and stellar corrugations separate or related phenomena?  (ii) What generates the ripples and how do they evolve?  (iii) More specifically, how old is the wave-like pattern and how long does it last? (iv) What can we learn from these disc corrugations? We attempt to find some answers in the present study. In Section 2, we summarise our minimalist galaxy model before presenting new analytic techniques in Section 3, and our final discussion in Section 4.

\section{A minimalist Galaxy model with gas} \label{s:model}

Our basic N-body framework for the Galaxy was introduced in our earlier work \citepalias{bla21e}, but here the model is extended to include a cold ($T = 10^3$ K), `light' ($M \approx 4 \times 10^9$  \Msun) gas disc. We refer to this extended model as the `hybrid' Galaxy model to distinguish it from our earlier, strictly `N-body' Galaxy model. A light gas disc is consistent with most earlier simulation work, although a threefold increase in the mass of cold gas has been adopted in a few studies \citep[see][]{bla16a}.

In our hybrid model, the Galaxy is approximated by a four-component system: 1) a dark matter (DM) host halo; 2) a central stellar bulge; 3) a thin stellar disc; 4) a cold gas disc. The relevant properties of each of these components (mass, structural parameters) are summarised in Tab.~\ref{t:comp}. This model is referred to as the `isolated hybrid' model.

To simulate the impulsive interaction between the Galaxy and an external perturber (a proxy for the Sagittarius dwarf; Sgr), we follow the approach used in our previous work. In brief, Sgr is approximated by a heavy ($M = 2 \times 10^{10}$ \Msun) point mass moving with high speed ($V \approx 330$~\kms) along a hyperbolic orbit that crosses the galactic plane at $R \approx 18$~kpc from the galaxy's centre. The simulation of the Galaxy-Sgr interaction is referred to as the `interaction hybrid' model, to distinguish it from our earlier `interaction N-body' model \citepalias{bla21e}.

\begin{table*}
\begin{center}
\caption{Galaxy model parameters. Columns 1 and 2 identify the galactic components and their associated functional forms; we note that these are approximations because they share the same gravitational potential. The total mass, scale length and cut-off radius are indicated in columns 3, 4, and 5 respectively. Column 6 is the number of collisionless particles used in the simulation (halo, bulge, disc) or to sample the gas disc distribution.
}
\label{t:comp}
\begin{tabular}{llccccc}
Component & Profile & Total mass & Radial scalelength & Cut-off radius & Particle count \\
 &  & $M_{\rm tot}$  & $r_s$ & $r_c$ & $N$ \\
 &  & ($10^{10}$ \Msun) & (kpc) & (kpc) & ($10^6$) \\
\hline
DM halo & NFW & 145 & 15 & 300 & 20 \\
Stellar bulge & Hernquist & 1.5 & 0.6 & 2.0 & 4.5 \\
Stellar disc & Exp, $\sech^2$ & 3.4 & 3.0 & 40 & 50\\
Gas disc & Exp, $\sech^2$ & 0.4 & 7.0 & -- & 2\\
\hline
\end{tabular}
\end{center}
\begin{list}{}{}
\item Notes: The NFW and Hernquist functions are defined elsewhere \citep[][]{nav97a,her90a}. The scaleheight of the stellar disc is $z_t \approx 250$ pc; the Toomre local instability parameter of the stellar disc is everywhere $Q \gtrsim 1.3$. The gas disc is isothermal with $T = 10^3$ K, with a scaleheight that varies with radius from roughly 20 pc near the centre to 160 pc at $R = 20$~kpc (a `flaring' disc). The gas disc is not truncated but merges smoothly with the background density (set at $10^{-20}$ \pcc\ in our \ramses\ setup).
\end{list}
\end{table*}

\subsection{Numerical experiment} \label{s:setup}

\subsubsection{Initial conditions}

The initial conditions (particle positions and velocities) of each of the components making up our Galaxy model are created with the Action-based GAlaxy Modelling Architecture software package \citep[\agama; ][]{vas19a}. We refer the reader to \citet[][their section 3]{tep21v} for a detailed description of \agama{}'s self-consistent modelling module for the collisionless components in our model (DM halo, bulge, stellar disc). 
We have complemented the \agama\ galaxy model framework to include the gas phase \citep[cf.][]{deg19a}; our methodology is described elsewhere (\textcolor{blue}{Tepper-García et al.} 2022b, in prep.). Here we provide only a brief description.

Our approach follows \citet[][]{wan10a} who give a prescription for setting up isothermal gas discs in equilibrium. In brief, the gas disc is isothermal and initially axisymmetric, with a surface density profile described by a radially declining exponential, $\propto \exp [ -R/R_g ]$, with a scalelength $R_g = 7$~kpc. Its vertical structure follows a $\sech^2[ z/z_0 ]$ profile, appropriate for a gas distribution in vertical hydrostatic equilibrium, with a scaleheight\footnote{We estimate the scaleheight by measuring the root-mean-square of the height of the gas with respect to the galactic plane as a function of $R$} $z_0$ that varies with cylindrical radius from  $z_0  \approx 20$ pc near the centre to $z_0 \approx 160$ pc at $R = 20$~kpc (a `flaring' disc). The radially increasing thickness of the gas disc is a consequence of the isothermal constraint and the fact that the (vertical) potential weakens with radius from the galaxy's centre. The azimuthal velocity profile of the gas disc ensures rotational support against radial instabilities \citep[see][their equation 13]{wan10a}.

\subsubsection{Evolution of the initial conditions} \label{s:sims}

We track the evolution of the following models: 1) the isolated hybrid model; and 2) the interaction hybrid model. The former is intended as a reference simulation to gauge the response of the synthetic galaxy to the interaction, and to assess the impact of any artefacts (e.g. numerical noise). In the interaction hybrid simulation, the point mass representing Sgr is placed at $\vec{r} \approx (-11.1, 0, 28.6)$~kpc with a velocity $\vec{v} = (-145.4,0,-220.2)$~\kms\ with respect to the galactic centre. 

For both simulations, the synthetic galaxy and the infalling point mass are evolved with the \ramses\ code \citep{tey02a} which incorporates adaptive mesh refinement (AMR). The system is placed into a cubic box spanning 600~kpc on a side. In addition to the galaxy and the perturber, the simulation box is initially filled with a cold ($T = 10^3$ K), tenuous, uniform gas. We do not attempt to set the ambient background gas in hydrostatic equilibrium, which is a non-trivial but feasible task \citep[e.g.][]{gro21g}. To deter the ambient gas from collapsing in a timescale shorter than our simulation time span, we set its density to an extreme threshold value ($10^{-20}$ \pcc).

We stress that our simulations are strictly isothermal, i.e. the gas is artificially kept at a temperature of $10^3$ K throughout. But to ensure this equation of state does not influence our dynamical results, we have run one additional (expensive) simulation to $t \approx 1$ Gyr with a different equation of state for the gas. This test simulation starts from the same initial conditions and using a setup identical to those of our isolated hybrid model, with the exception of the following: 1) The ambient gas is initially hot ($T = 10^6$ K); 2) we include an idealised, uniform ultraviolet (UV) background (appropriate for a redshift $z = 0$) that heats the gas; 3) we allow the gas to cool radiatively, assuming it is composed of hydrogen, helium, and a mixture of metals resulting in a metallicity equivalent to 10 percent of the solar value. To deter the gas from cooling indefinitely, we set a temperature floor at $T = 10^3$ K. In addition to the effect of UV photo-heating, the gas temperature increases as a result of compression.

We have compared the evolution of the discs with cooling/heating and the strict isothermal condition, and while there are differences between them, none are important for our dynamical study. This may seem contradictory to our assertion that the setup of the gas disc must be strictly isothermal. However, due to the existence of a temperature floor, the high gas density across much of the gas disc as well as its relatively high metallicity, the gas disc to a large extent retains its equilibrium temperature (and thus its structure) throughout, while the ambient medium barely cools during the time span of the simulation. In other words, the simulation starts and ends up (after $\sim 2$ Gyr) with virtually two distinct gas phases: a cold, dense disc, and a hot, tenuous background, connected by an insignificant intermediate temperature phase.

Given the similarity between the simulation with cooling/heating and the isothermal condition, and also because the former is significantly more expensive to run, we choose to investigate the corrugations in stars and gas with our isothermal simulations.

In addition to neglecting any cooling/heating processes, we ignore the presence of a hot halo around the Galaxy \citep[e.g.][]{mil13a}, or the presence of magnetic fields \citep[e.g.][]{jan12a}. We also neglect galactic processes such as star formation or stellar feedback of any kind. Our aim is to construct the simplest possible model to isolate the dynamical effect of the interaction from other, potentially obfuscating processes. More realistic simulations incorporating the physics we have neglected will be explored in future studies.

The total simulation time is about 2 Gyr for both the isolated hybrid model and the interaction hybrid model. At runtime, the AMR grid is maximally refined up to level 14, implying a limiting spatial resolution of \mbox{600~kpc / $2^{14}  \approx 36$ pc}. As a result, the vertical structure of the gas disc is barely resolved at $R \lesssim 4$~kpc.

A requirement of numerical experiments aimed at investigating the response of a system to a prescribed stimulus is for the system to be in near-perfect equilibrium when left in isolation, and for it to maintain that state for a time span that is long compared to the actual experiment. We have shown in our previous work that the isolated N-body model preserves its initial state for at least 4 Gyr \citepalias{bla21e}. Since the isolated hybrid model follows virtually an identical setup, we expect this model to retain the same degree of stability. To this end, we analyse the overall stability of isolated hybrid model by comparing the initial structural and kinematic properties of the stellar disc and of the gas disc with its state after 2 billion years of evolution. A detailed discussion of this analysis is deferred to the Appendix. Here it suffices to say that the synthetic galaxy, and in particular both the stellar and gas discs, maintain their initial state throughout their evolution in isolation, implying that our setup yields a stable model from the outset.\\

To summarise: In what follows, we discuss three different simulations: 1) isolated hybrid simulation; 2) interaction hybrid simulation; and 3) interaction N-body simulation. Simulations 1 and 2 are new and are presented for the first time in the present paper. Simulation 3 was introduced and discussed at length in \citetalias{bla21e}.

\section{Results} \label{s:result}

We focus on the evolution of the vertical response of the stellar and gas discs triggered by the one-time interaction with the crossing satellite. In order to accurately measure the vertical displacement and velocity of the stars and gas at any given point across the disc, and for consistency across time steps, we need to calibrate each snapshot. We do so by correcting for the shift and velocity of the centre of mass (CoM) of the stellar disc induced by the interaction. In order to compensate for a potential offset in the CoM as a result of distortions in the outer disc, we calculate the CoM in an iterative way by considering only the stellar mass within a prescribed radius (20 kpc) that is decreased by half after each iteration until convergence \citep[a.k.a. the `shrinking sphere' approach;][]{pow03o}. The latter is achieved when the change in the CoM distance between consecutive iterations is less than a specified tolerance (0.05 kpc). After the entire system is shifted to the position and velocity of the CoM, we correct for the inclination of the galaxy's midplane by calculating the angular momentum vector $\vec{L}$ of the stellar disc considering only the stars within the last radius attained when calculating the CoM, and by rotating the entire system such that $L_z$ (the $z$-component of $\vec{L}$) is parallel to the $z$-axis of the simulation box.

At each epoch during the course of each simulation, we  estimate the average vertical displacement,
\mbox{$
	\overline{z} \equiv \langle z \rangle
$}
, and the average vertical velocity,
\mbox{$
	\overline{V}_z \equiv \langle V_z \rangle
$}
, separately for the stars in the disc and for the gas disc in the following way. First, we consider a face-on projection of the synthetic galaxy over a fixed spatial range, in this case \mbox{ $[-20, 20] \times [-20, 20] ~\kpc^2$}, and divide this projection into a rectangular mesh with $N = 500$ equal bins (or pixels) per side, (yielding a total of $M = N^2$ bins). We then look at a patch of gas or stars in a bin (40 pc in size) centred around the point $p(x,y)$ on either projection, and calculate the {\em median} $z$- or $V_z$-value in each bin, separately for the gas and for the stars.

\begin{figure}
\centering
\includegraphics[width=0.47\textwidth]{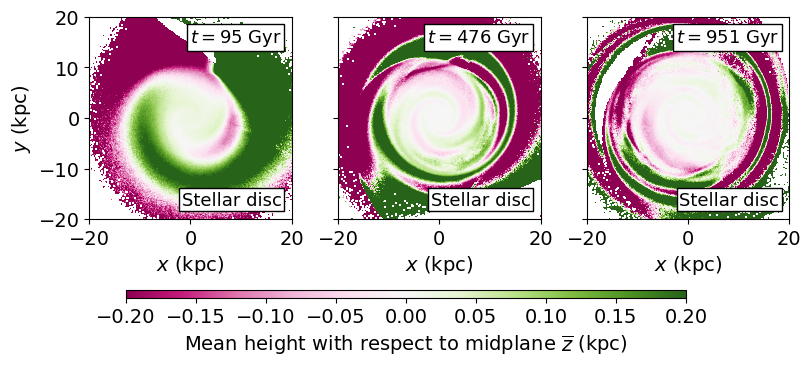}
\includegraphics[width=0.47\textwidth]{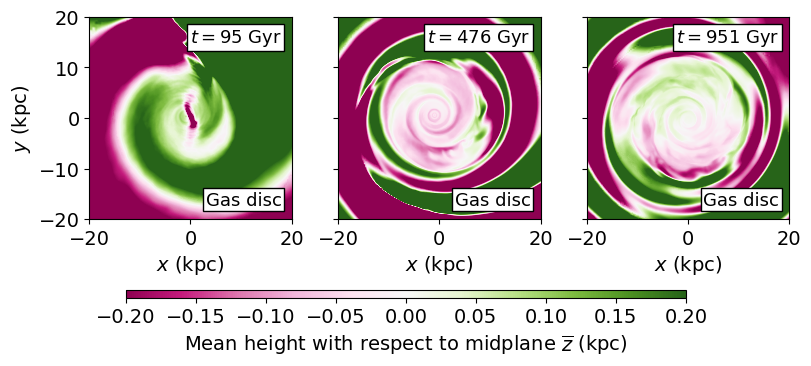}
\caption[  ]{ Mean height (in kpc) with respect to the galactic midplane of the stars in the disc (top) and gas (bottom) $\sim 100$ Myr after the impact (left), $\sim 380$ Myr after the impact (middle), and $\sim 850$ Myr after the impact (right) in the interaction hybrid simulation. }
\label{f:z_mean}
\end{figure}

\begin{figure}
\centering
\includegraphics[width=0.47\textwidth]{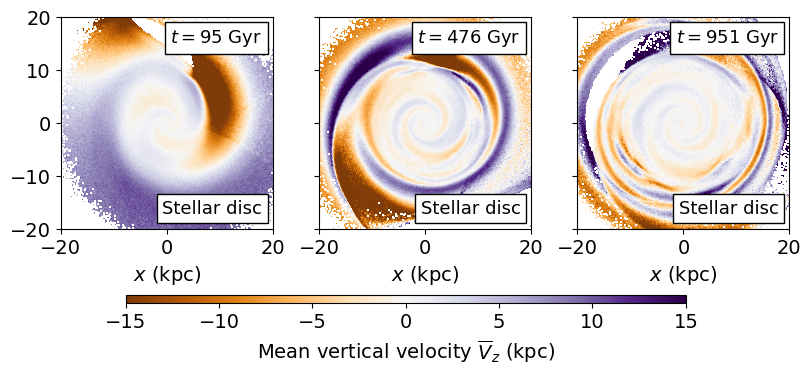}
\includegraphics[width=0.47\textwidth]{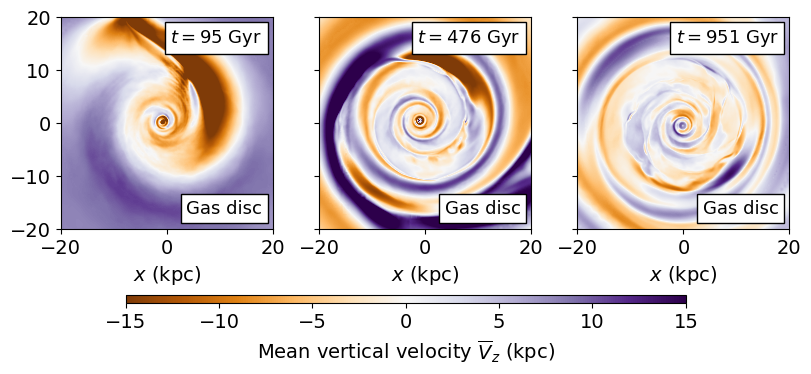}
\caption[  ]{ Mean vertical velocity of the stars in the disc (top) and gas (bottom) $\sim 100$ Myr after the impact (left), $\sim 380$ Myr after the impact (middle), and $\sim 850$ Myr after the impact (right) in the interaction hybrid simulation.}
\label{f:vz_mean}
\end{figure}

\subsection{Interaction and overall evolution of the stellar and gaseous discs} \label{s:inter}

Roughly $100$ Myr after the start of the interaction hybrid simulation, the point mass representing the satellite reaches the galactic midplane, and it crosses the stellar and the gas discs at $R \approx 18$~kpc, triggering a violent response in both. In line with our previous pure N-body interaction simulation, the stellar disc develops a one-sided ($m = 1$) perturbation that quickly develops into a bisymmetric ($m = 2$) bending wave (Figs.~\ref{f:z_mean} and \ref{f:vz_mean}, top rows; see also \citetalias{bla21e}, their figs. 6 - 8). In addition, the interaction triggers a kinematic density wave that eventually leads to the formation of a two-arm spiral structure.

A visual inspection of the mean vertical displacement $\overline{z}$ of the gas (Fig. ~\ref{f:z_mean}, bottom row) reveals that it is, at least qualitatively, very similar to that of the stars in the disc (Fig. ~\ref{f:z_mean}, top row). The same is true for the vertical velocity $\overline{V_z}$ (compare the top and the bottom rows of Fig. \ref{f:vz_mean}). Thus, apparently the gas disc mimics the behaviour of the stellar disc, both spatially and kinematically, throughout its evolution out to $\sim 500$ Myr (equivalent to roughly two rotation periods at the solar circle after impact), in qualitative agreement with \citet{gom17a}. Therefore, we conclude that `the gas follows the stars.' We caution the reader, however, the time scales involved are rather short, and that this behaviour may not hold on the long term ($\gtrsim 1$ Gyr), with exception perhaps of the gas and stars in the outer disc \citep[e.g.][]{lap18a}.

It is also apparent that the disc response is stronger with increasing radius \citep[][]{ede97b,pog21f}. The latter is not surprising given that the vertical restoring force in the vicinity of the Galactic midplane declines with radius.

\subsection{Corrugations in stars and gas}

\subsubsection{Phase correlation} \label{s:pcorr}

While the stellar and gas discs are fundamentally different in character, we have already seen that they share a common evolution at the outset but evolve separately at later times. To understand the mechanism behind this divergent behaviour, we explore whether there exists a measurable correlation between the vertical response of the stellar and gas discs to the perturbation induced by the satellite.
We consider a quantitative analysis of $z$ and $V_z$ {\em simultaneously} by mapping the spatial and the kinematic data to the vertical phase space $(z,V_z)$.  Consider a pair of $\overline{z}$ and $\overline{V_z}$ maps of the synthetic galaxy at a given epoch (Figs. ~\ref{f:z_mean} and  ~\ref{f:vz_mean}). But instead of looking at $\overline{z}$ or $\overline{V_z}$ separately, we combine these two quantities into a vector, an element of $(z,V_z)$ space. However, because $z$ and $V_z$ carry different units, we apply a scale transformation to bring them to a common dimension. The natural choice is to scale $\overline{z}$ by the vertical frequency $\nu$, defined by
\mbox{$
	\nu^2(R) = d^2 \Phi / d z^2 \, ,
$}
where $\Phi(R,z)$ is the total potential at $(R,z)$, and the derivative is evaluated at $z = 0$.

\begin{figure}
\centering
\includegraphics[width=0.4\textwidth]{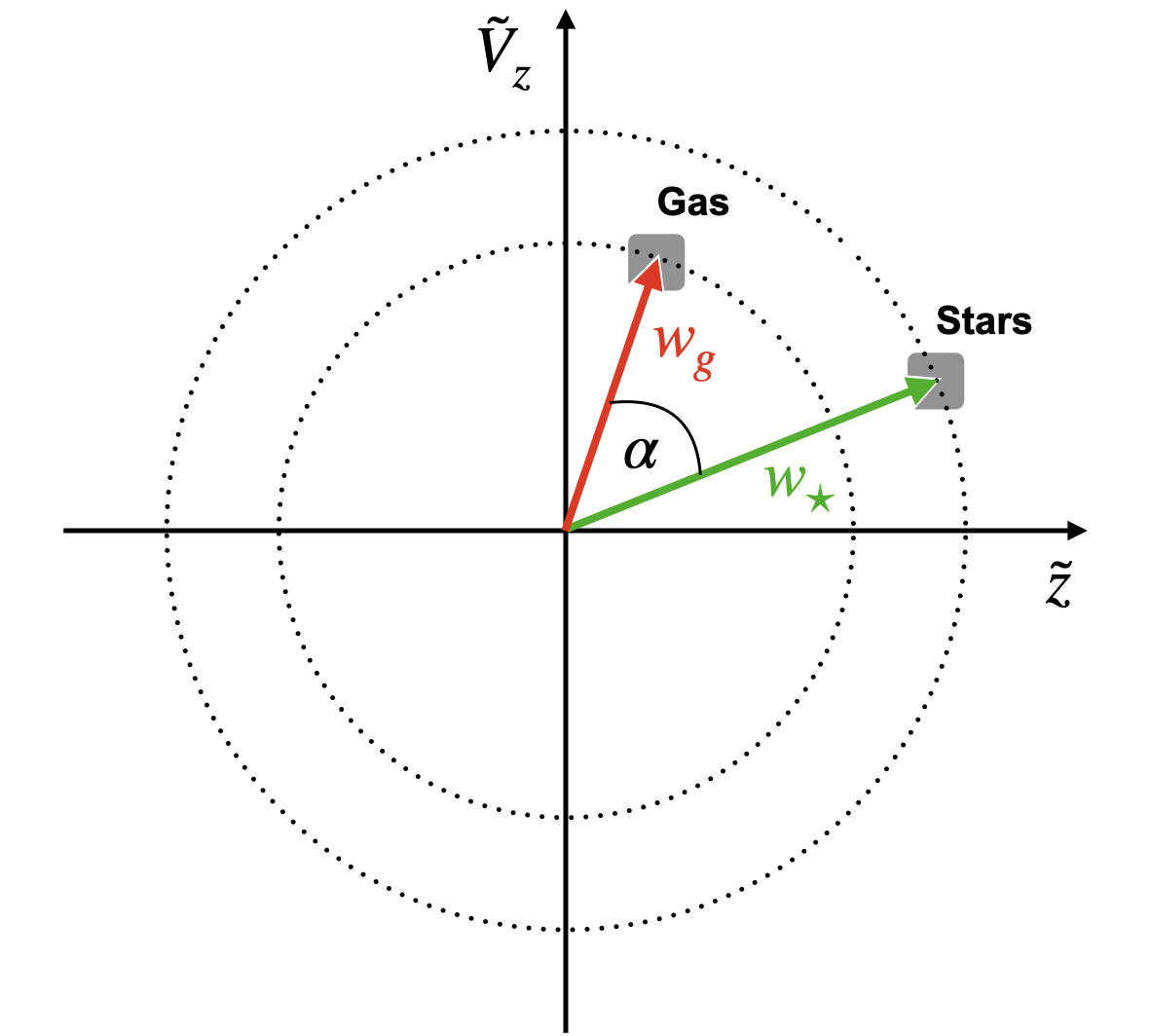}\\
\vspace{-5pt}
\caption[  ]{  Schematic representation of the location in vertical phase-space of an ensemble of stars (pointed at by the green vector) and a patch of gas (pointed at by the red vector) found on a point $(x,y)$ in the galaxy disc, and the angle (phase) difference, $\alpha$, between these. Note that the phase space coordinates $(\tilde{z}, \tilde{V}_z)$ are rescaled versions of the counterparts $(z, V_z)$ (Eq.~\ref{e:trafo}). }
\label{f:phase_space}
\end{figure}

Thus we define the normalised phase space coordinates
\begin{equation} \label{e:trafo}
	\tilde{z} \equiv \frac{ 1 }{ \sqrt{2} } ~\nu ~\overline{z}   \quad ; \quad \tilde{V}_z \equiv \frac{ 1 }{ \sqrt{2} } ~\overline{V}_z
\end{equation}
and the corresponding phase-space vectors,
\mbox{$
	w_g \equiv \{ \tilde{z}, \tilde{V}_z \}_g
$}
and
\mbox{$
	w_\star \equiv \{ \tilde{z}, \tilde{V}_z \}_\star \, ,
$}
where the subscript is used to distinguish between the vectors corresponding either to the gas or to the stars (Fig.~\ref{f:phase_space}). These vectors are elements of the plane spanned by $(\tilde{z}, \tilde{V}_z)$, a transformed version of the vertical phase space. Note that this transformation renders the distribution of stars on this plane  roughly circular, regardless of their location in cylindrical radius $R$.

Within the transformed phase space, we measure the angle $\alpha$ between the vectors $w$ geometrically via
\begin{equation} \label{e:phase}
	\cos \alpha = \frac{ w_g \cdot w_\star }{ |w_g| ~|w_\star| }  \,.
\end{equation}
where $|w|$ indicates the length of the vector $w$ to the centroid of the distribution.

\begin{figure*}
\centering
\includegraphics[width=0.32\textwidth]{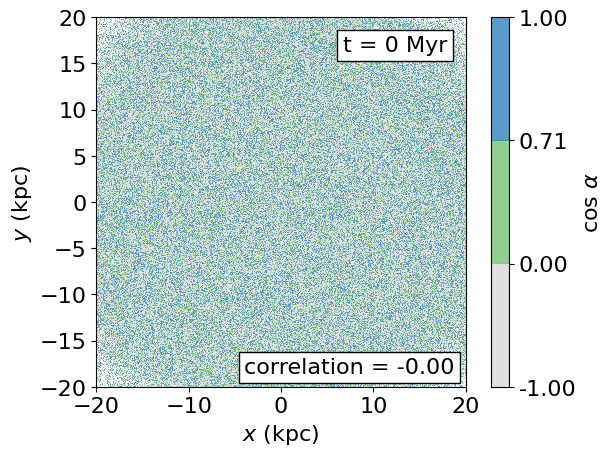}
\includegraphics[width=0.32\textwidth]{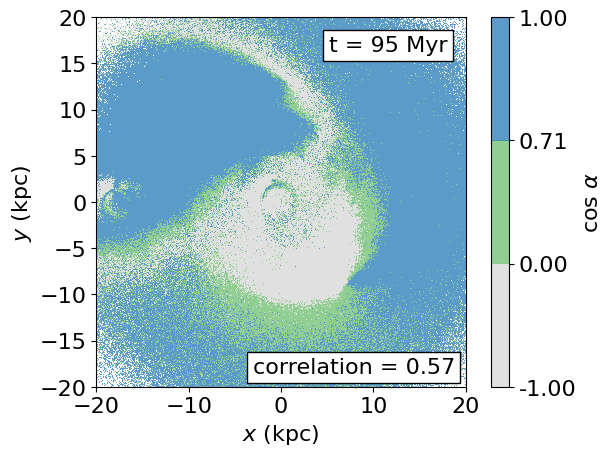}
\includegraphics[width=0.32\textwidth]{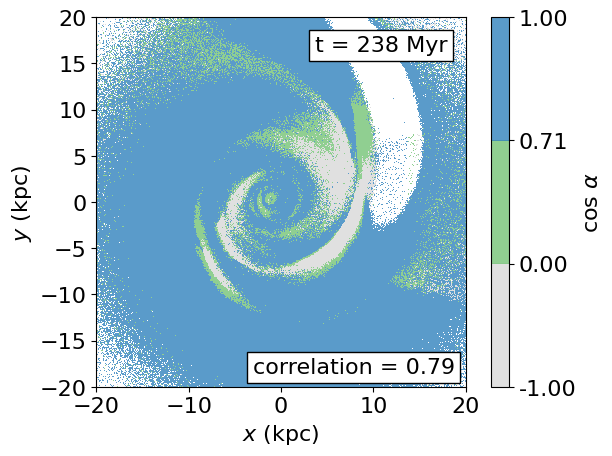}
\includegraphics[width=0.32\textwidth]{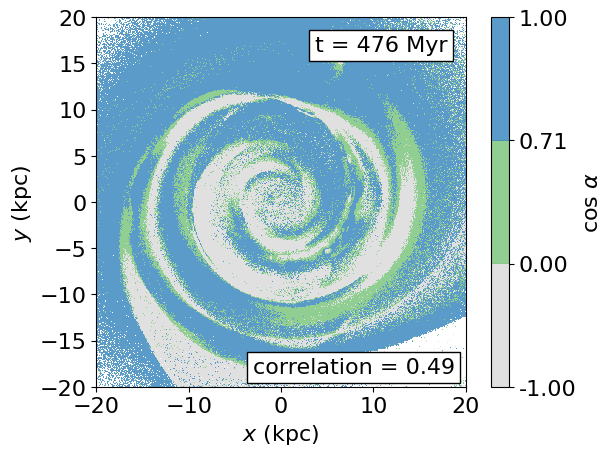}
\includegraphics[width=0.32\textwidth]{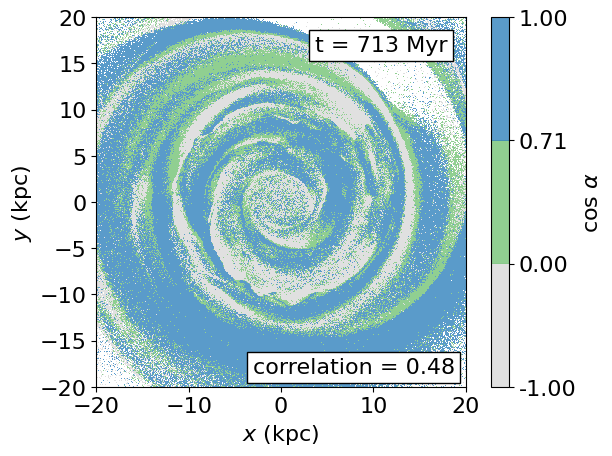}
\includegraphics[width=0.32\textwidth]{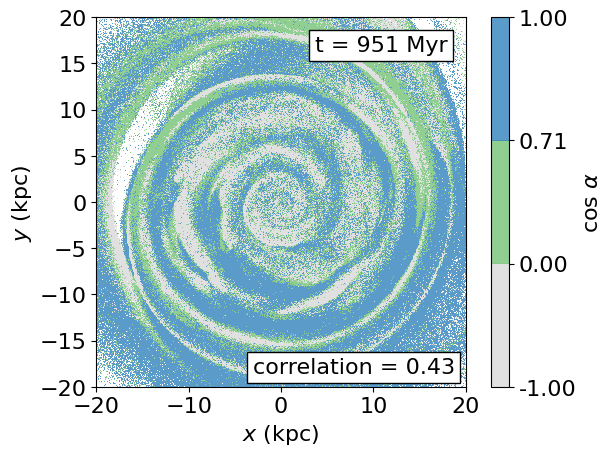}
\caption[  ]{ Evolution of the angle $\alpha$ between stars and gas (Eq.~\ref{e:phase}) at selected epochs from roughly -100 Myr prior to the impact to roughly 850 Myr after the impact in the interaction hybrid simulation. Each panel and each displays the cosine of $\alpha$ between the phase space vector of an ensemble of stars and a patch of gas around each location $p(x,y)$ (see Fig.~\ref{f:phase_space}). Highly correlated corrugations are identified by a uniform blue colour, while weakly correlated corrugations are identified by a uniform green colour. In contrast, a uniform grey colour, in contrast, indicates non-correlated or anti-correlated star-gas corrugations (see text for details). Each panel corresponds to a given epoch as indicated by the legend on the top-right corner. The total correlation $\mathcal{C}$ between the vertical oscillations of stars and gas across the galaxy, measured by the sum of all $\cos \alpha$ in that panel, is indicated on the bottom-right corner (Eq.~\ref{e:corr}). The impact site at impact time $t = 95$ Myr  is clearly visible at $\vec{r} \approx (-18, 0)$~kpc (top row, central panel)
}
\label{f:gas_star_phase_corr}
\end{figure*}

Evaluation of Eq.~\eqref{e:phase} over a pair of maps such as the left panels of Figs.~\ref{f:z_mean} and \ref{f:vz_mean} yields values in the range $[-1,1]$; non-negative values correspond to angles (measured in radian) in the range \mbox{$ | \alpha | \leq \pi / 2$}, and thus indicate that stars and gas at $p(x,y)$ oscillate vertically {\em in phase}. Negative values indicate the opposite, i.e. the gas and stars are {\em out of phase}. We further categorise the in-phase corrugations in stars and gas to be {\em strongly} correlated if \mbox{$| \alpha | \leq \pi / 4$} (corresponding to \mbox{$\cos \alpha \gtrsim 0.71$}) or {\em weakly} correlated if \mbox{$\pi / 4 \leq | \alpha | \leq \pi / 2$} (corresponding to \mbox{$0 \leq \cos \alpha \lesssim 0.71$}).

Using the approach just described, we calculate the angle $\alpha$ and the correlation at each time step ($\Delta t \approx 10$ Myr) in our simulation from $t = 0$ Gyr to $t \approx 2$ Gyr, i.e. from roughly -100 Myr prior to the impact to roughly 1800 Myr after the impact. Thus we are able to quantify the evolution of the correlation angle $\alpha$ between the vertical motion of stars and gas across the galaxy. In order to quantify the occurrence of by-chance correlations, we repeat the analysis, but this time rotating each of the $\overline{z}_\star$ and  $\overline{V_z}_\star$ maps anti-clockwise by $\pi/2$ prior to calculating the vectors $w_\star$.  We refer to the latter as our `control data', and to the former as our `simulation data'.

We further quantify the correlation {\em globally} (at each time step) by calculating the arithmetic average of the pixel-wise values delivered by Eq.~\eqref{e:phase},
\begin{equation} \label{e:corr}
	\mathcal{C} \equiv \langle \cos \alpha \rangle = \frac{ 1 }{ M }\sum_{i=1}^{M} \left( \cos \alpha \right)_i
\end{equation}
where $\left( \cos \alpha \right)_i$ is the value of the angle $\alpha$ at pixel (or bin) $i$, and $M$ is the total number of bins in the map (see beginning of Sec.~\ref{s:result}). Analogous to the angle $\alpha$, positive values of $\mathcal{C}$ indicate a positive correlation between stars and gas across the map, while negative values indicate the opposite.

\begin{figure}
\centering
\includegraphics[width=0.48\textwidth]{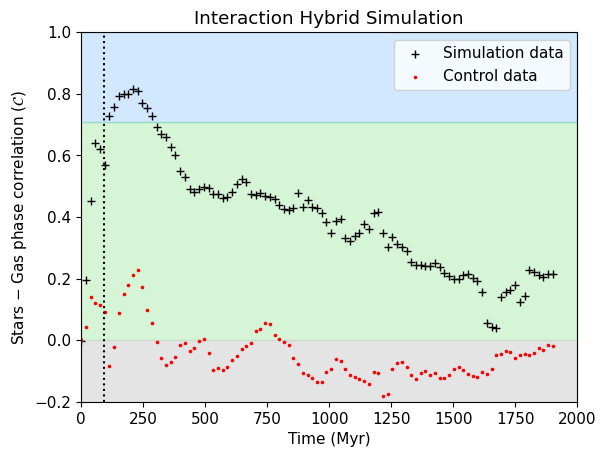}
\includegraphics[width=0.48\textwidth]{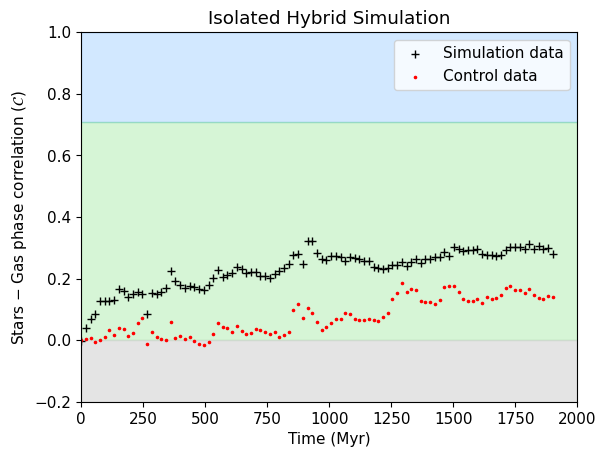}
\caption[]{  The evolution of the phase correlation $\mathcal{C}$ of the joint stellar and gas corrugations  (Eq.~\ref{e:corr}) from roughly -100 Myr prior to the impact (tagged by the vertical dotted line) to roughly 1800 Myr  after the impact is shown in the bottom-right panel (black `$+$' symbols). Top: Interaction hybrid simulation. Bottom: Isolated hybrid simulation. In both panels the red dots indicate the evolution of $\mathcal{C}$ for our control data. The blue / green / grey shaded areas follow the colour coding of Fig.~\ref{f:gas_star_phase_corr}  (see text for details). }
\label{f:gas_star_phase_corr_evol}
\end{figure}

The results of our approach are displayed in Fig.~\ref{f:gas_star_phase_corr}. There we present a series of panels, each corresponding to a face-on projection of the synthetic galaxy at a different epoch in the system's evolution from its initial state roughly 100 Myr prior to impact (top left), to about 850 Myr after the impact (bottom right). The colour coding indicates the value of $\cos \alpha$: blue and green correspond to \mbox{$\cos \alpha \gtrsim 0.71$} and \mbox{$0 \leq \cos \alpha \lesssim 0.71$}, i.e. to strong and weak in-phase oscillations, respectively, while a grey colour corresponds to negative values, and thus indicate an out of phase oscillation. In each panel, the simulation time is indicated in the top right corner, and the value of $\mathcal{C}$ in the bottom right corner. 

Initially, i.e. prior to the impact (top left panel), there is no correlation between the vertical response of the stars and the gas because an equilibrium distribution has no centroid shift in the transformed phase space. 

At impact time ($t = 95$ Myr, central panel, top row), we see a signal consistent with a strong in-phase vertical response between stars and gas at $R \gtrsim 5$~kpc. The correlation index increases up to about 150 Myr after impact; its value climbs from roughly 0.6 to 0.8 (top row, right panel). After that, the oscillations in the stars and gas start to move out of phase, as indicated by a weakening of the overall signal (bottom row), a trend that continues until the end of the simulation.

It is interesting that the difference in correlation angle between stars and gas reveals a great deal of substructure in the disc, notably the spiral structure. Presumably, the contrast between stars and gas in this space arises because the gas is partially compressed by the stellar density wave, and is thus displaced downstream, causing it to move out of phase with respect to the stars along the density wave and generating a `galactic shock' \citep[][]{fuj68m,rob69r,lub86s,bab15t}.

The global behaviour of the phase correlation is captured in the top panel of Fig.~\ref{f:gas_star_phase_corr_evol}. The black symbols ({\bf +})
trace the behaviour of the global correlation angle between the corrugations of the stellar disc and the gas disc (Eq.~\ref{e:corr}) over the full simulation. The red symbols (\textcolor{red}{$\bullet$}) indicate the corresponding result for our control data. The background has been colour-coded following the same convention followed in Fig.~\ref{f:gas_star_phase_corr}. The vertical dotted line flags the impact epoch. In this form, it becomes apparent that the corrugations of stars and the gas in the disc are strongly correlated ($\mathcal{C} \gtrsim 0.71$, blue shaded area) for roughly 150 Myr after the impact, and only weakly correlated after that ($0 \lesssim \mathcal{C} \lesssim 0.71$, green shaded area), consistent with the results presented in Fig.~\ref{f:gas_star_phase_corr}. It is worth noting that the level of correlation reached at $t \approx 1.3$ Gyr is comparable to the level reached at the end of the isolated hybrid simulation, likely seeded by numerical noise (see Fig.~\ref{f:gas_star_phase_corr_evol}, bottom panel), and even drops below this level at later times. This is consistent with the dynamical response of the discs triggered by the interaction dominating over numerical noise effects for a significant fraction (if not all) of the simulation time span. 

Our findings suggest that, in the absence of any further, strong perturbations, estimates of the degree of correlation between the corrugations observed in stars and gas can be used to age-date the phenomenon, regardless of whether the perturbation is extrinsic \citep[][]{elm95t} or intrinsic \citep[][]{wei91a} to the galaxy, provided the origin of the perturbation does not lie more than $\sim 1$ Gyr in the past.

\subsubsection{Vertical spatial and kinematic amplitudes} \label{s:amp}

Our impulsive simulation provides the ideal framework to study the vertical response of the stellar and gaseous discs to a one-time interaction with an external perturber \citep[cf.][]{cha09y}, without the obfuscating effects of subsequent disc crossings \citep[e.g.][]{che18a,lap18b}.

In order to analyse the vertical response of the stellar and gaseous discs throughout their evolution, we proceed as explained next. We stress that our approach is preferred over using Fourier methods because we find a full modal analysis is not warranted here, at least not in our initial analysis.

\begin{figure*}
\centering
\includegraphics[width=.32\textwidth]{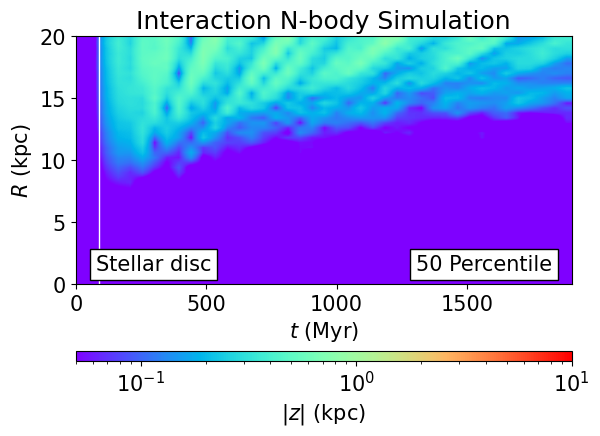}
\includegraphics[width=.32\textwidth]{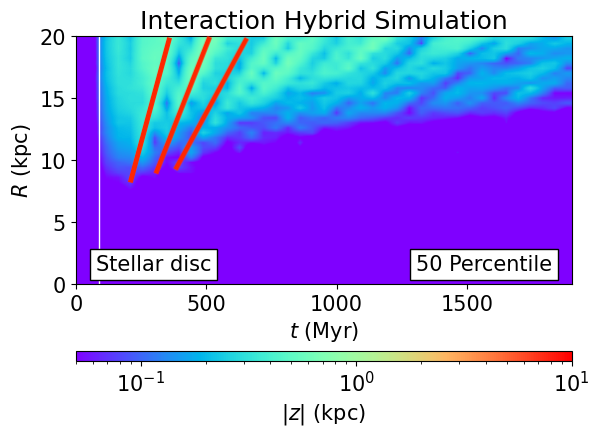}
\includegraphics[width=.32\textwidth]{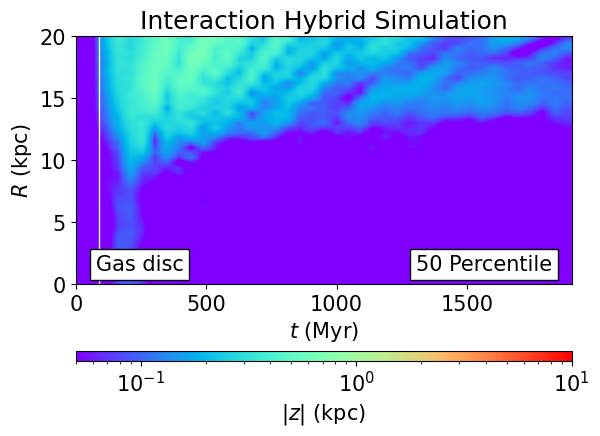}
\includegraphics[width=.32\textwidth]{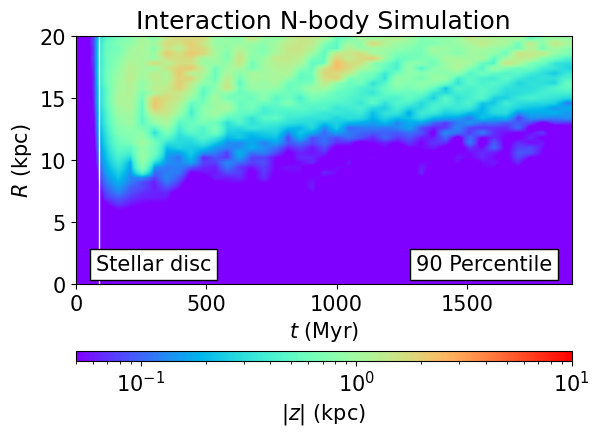}
\includegraphics[width=.32\textwidth]{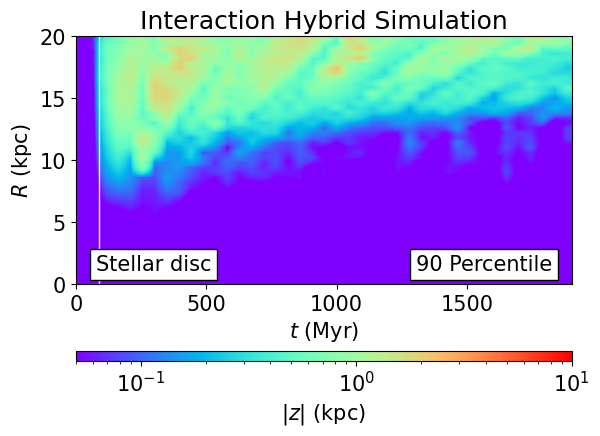}
\includegraphics[width=.32\textwidth]{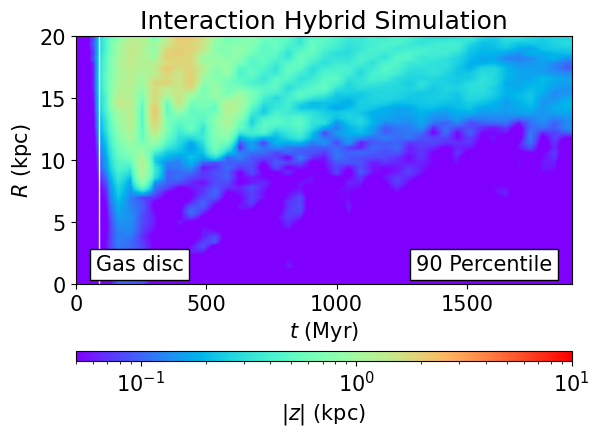}
\includegraphics[width=.32\textwidth]{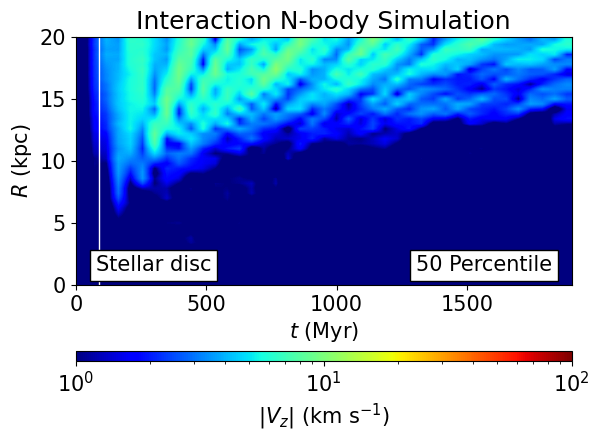}
\includegraphics[width=.32\textwidth]{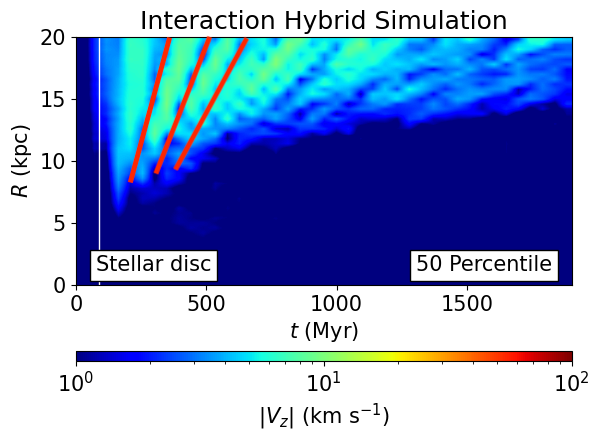}
\includegraphics[width=.32\textwidth]{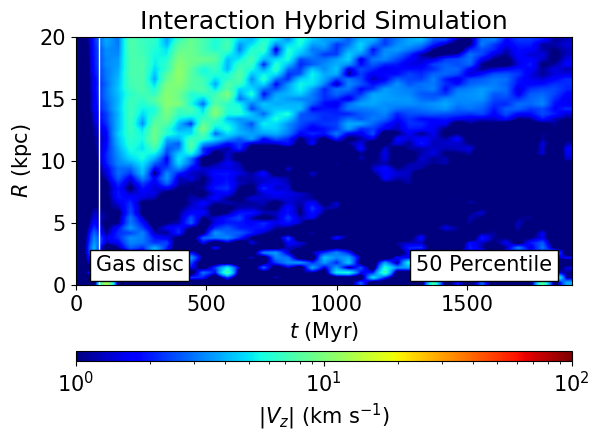}
\includegraphics[width=.32\textwidth]{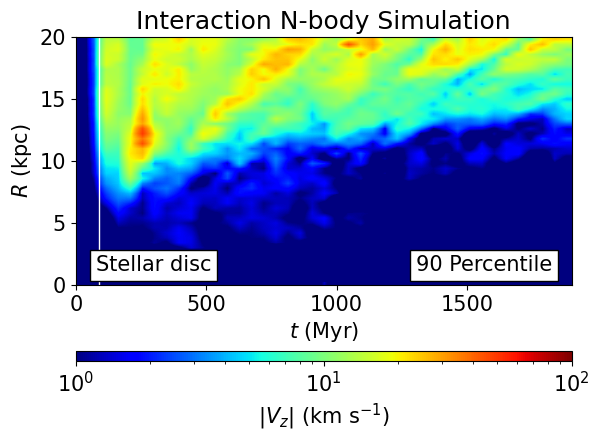}
\includegraphics[width=.32\textwidth]{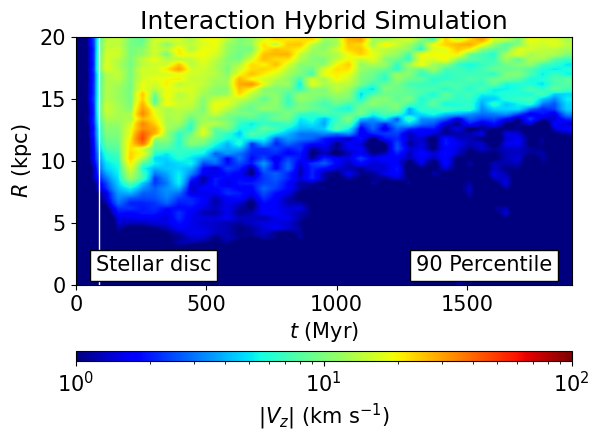}
\includegraphics[width=.32\textwidth]{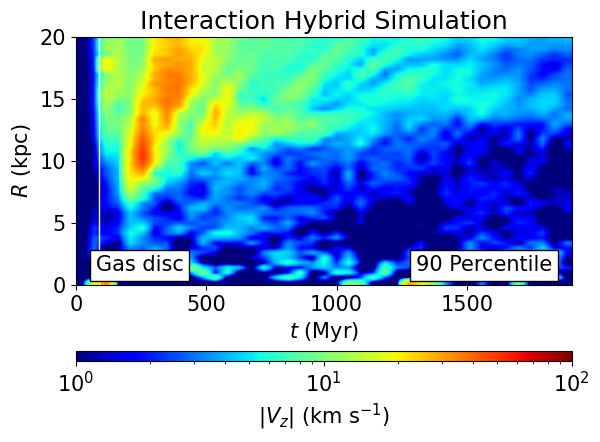}
\caption[]{ Spatial and kinematic amplitude of the corrugations. Each panel displays the evolution in time of the absolute value of either the vertical displacement $| z |$ (in kpc; rows 1 and 2) or the vertical velocity $| V_z |$ (in \kms; rows 3 and 4) across the disc at each radius. The first and third rows display the amplitude in $| z |$ or $| V_z |$ at the 50 percentile level, respectively, while the second and fourth rows display these quantities at the 90 percentile level. The left column displays the results for the stellar disc in interaction N-body model, while the central column and right columns corresponds to the interaction hybrid model. The central column displays the result for the stellar disc, and the last column, the result for the gas disc. The white vertical line flags the epoch of the impact at $t \approx 100$ Myr and it is identical in all panels. The red lines in the rows 1 and 3 (central column) are identical and illustrate the fact that maxima in the vertical displacement are coincident with minima in the vertical velocity (and vice-versa). See text for further details. }
\label{f:gas_star_amplitude}
\end{figure*}

At each time step during the evolution of the interaction hybrid model, the absolute value of $z$ and $V_z$  is mapped from the Cartesian $(x, y)$ plane onto the polar $(R, \phi)$ plane, where $R$ and $\phi$ are the cylindrical radius and the azimuthal angle, respectively. We do the same for the isolated hybrid model, which we then use to correct\footnote{In practice, we subtract one map from the other before taking the absolute value; our results are virtually identical if we skip this step. } each $(R, \phi)$ map (i.e. at each time step) for any vertical displacement or velocity that results from numerical noise. Then, for each corrected $(R, \phi)$ map, we calculate the 50 percentile (median) and 90 percentile (here considered a proxy for the maximum value) at each radial bin {\em along the full range in azimuth}. Thus we obtain an estimate of the median and maximum vertical displacement and vertical velocity at each radius $R$ across the disc, at each time $T$, yielding two $|z|$ maps and two $|V_z|$ maps, one for the stellar disc and one for the gas disc, on the $(T,R)$ plane. Furthermore, to assess the potential effect of the gas disc on the stellar disc, we repeat the analysis, this time for the stellar disc in the interaction N-body model. The results are displayed in Fig.~\ref{f:gas_star_amplitude}.

The first two rows display the $|z|$ maps, the next two rows the $|V_z|$ maps. The first and third rows display the 50 percentile in $| z |$ and $| V_z |$, respectively, while the second and fourth rows the 90 percentile of each quantity. The left column displays the results for the stellar disc in interaction N-body model, while the central column and right columns corresponds to the interaction hybrid model; the central column displays the result for the stellar disc, and the last column, the result for the gas disc.

Prior to the impact (whose epoch is flagged by the white vertical line at $t \approx 100$ Myr), neither the stellar disc (in either simulation) nor the gas disc reveal any significant vertical response, consistent with the fact that up to this point the synthetic galaxy can be regarded as an isolated system. But the situation changes dramatically after the interaction: clearly, both the stellar and the gas disc undergo a strong vertical perturbation, a bending wave.

The perturbation, as revealed either by the change in $| z |$ or $| V_z |$, is apparently structured, displaying a striation pattern with ridges running from top to bottom that become more slanted with time. The ridges correspond to small $| z |$ or small $| V_z |$ amplitudes, and are thus associated with the `nodes' of the bending wave, i.e. the regions across the disc where the bending wave intersects the midplane. Their inclination (with respect to, say, the $R$-axis) on the $(T,R)$ plane is directly related to the degree of windup of both the kinematic density wave (`spiral arms') and the bending wave; the higher the degree of wind-up, the more inclined are these ridges.\footnote{ A near-perfectly circular and stationary bending wave node -- an example of extreme windup -- corresponds to a near-perfectly horizontal ridge in the $(T,R)$ plane. }

Apparently, the perturbation induced by the satellite does not cover all radii, but extends only from $R = 20$~kpc into $R \approx 7 - 9$~kpc, leading to the formation of a lower `envelope' on the $(T,R)$ plane. This is a natural consequence of the fact that the restoring force of the disc, or equivalently, its resilience to any external perturbation, increases towards the centre \citep[cf.][their figure 17]{lap18b}. Moreover, the width of the swath along $R$ on the $(T,R)$ plane decreases with time, and does so in an in-out fashion, implying that the coherent vertical motion of the stars, and of the gas, washes away with time, in agreement with \citet[][]{pog21f}. We will return to this point later in Sec.\ref{s:ener}.

We now look at some important differences across panels, i.e. differences between the spatial and kinematic responses of a stellar disc in the absence or presence of a gas disc, and the difference between in the spatial and kinematic responses of a gas disc compared to a stellar disc.

We note first that the response in the vertical displacement as given by $| z |$ is offset by $\pi / 2$ from the vertical velocity, i.e. maxima in the vertical displacement at any radius and any time are coincident with minima in the vertical velocity, and vice-versa. To illustrate this point, we include in the panel, at the centre of the top row, a series of red lines roughly tracing the first four minima in the vertical displacement (50 percentile) of the stellar disc in the interaction hybrid model. We insert an identical set of lines in the central panel of the third row. Clearly, these lines trace the maxima in $| V_z |$. The same is true at the 90 percentile level, and it is also true for the gas disc. This result is not trivial and indicates that the corrugations induced by the satellite in both the stellar and the gaseous discs behave predominantly as travelling ripple patterns, initially in phase before drifting out of phase, as discussed earlier \citep[cf.][for the same effect in a pure N-body simulation]{pog21f}. This interpretation of Fig.~\ref{f:gas_star_amplitude} justifies our earlier method in using a correlation angle as a way of tracking the relative evolution of the joint star-gas corrugations (Sec.~\ref{s:pcorr}).

We now focus on the stellar disc in both the interaction N-body simulation and the interaction hybrid simulation (left and central columns). It is reassuring that the behaviour of the stellar disc in these two simulations is very similar. This is true both for $| z |$ (rows 1 and 2) and $| V_z |$ (rows 3 and 4), and both at the 50 percentile level (rows 1 and 3) and at the 90 percentile level (rows 2 and 4). The response in each case are obviously not identical, but the differences appear to be minor. We note that the comparison between these two simulations is justified, as the stellar discs in them are virtually identical to one another by design (see Fig.~\ref{f:surfdens_prof}).

The similarity between the vertical response of the stellar discs in these different simulations indicates that a light, cold gas disc as we have included in our hybrid model does not significantly affect the dynamical behaviour of the stellar disc. This is important because it suggests that the onset and development of the phase spiral detected by Gaia is not strongly influenced by the presence of gas. We address this issue in a future paper.

We now compare the vertical response of the stellar and gaseous discs in the interaction hybrid simulation (central and right columns). Within the first 700 Myr after the impact, the two components appear to respond in a similar way to the perturbation, both in terms of $| z |$ (rows 1 and 2) and $| V_z |$ (rows 3 and 4). The median vertical displacement is $ | z | \approx 1$~kpc and the median vertical velocity $| V_z | \approx10$~\kms. At the 90 percentile level, the vertical displacement reaches values of up a few kpc, while the corresponding vertical velocity may reach values of several tens of \kms. These ranges are in good agreement with the corresponding values observed in the Galaxy both in the stellar disc \citep{wid12a} or the gas layer \citep{alv20a}, or in the gaseous layers of external galaxies \citep{san15h,nar20a}. 

After $t \approx 700$ Myr, however, we observe a significant difference between the ongoing response of the stellar disc and that of the gas disc. Indeed, while both discs appear to settle with time, the gas disc does so more quickly, as indicated by the fading amplitudes in $| z |$ and $| V_z |$ and both at the 50 percentile and 90 percentile levels. This fact is consistent with, and may actually explain, the decrease in the correlation between the corrugations in stars and gas discussed in Sec.~\ref{s:pcorr}. Likely, the gas waves dampen at a faster rate compared to the stellar waves because the gas is subject to compression, which transforms kinetic energy into internal energy that is radiated away in order to satisfy the isothermal constraint imposed on the gas. 

We note finally the near-vertical line apparent in the maps displaying to the response of the gas disc (right column) at around $t \approx 200$ Myr. This corresponds to the external perturber along its polar orbit and which -- as it crossed the disc -- accreted gas and thus becomes visible in this plane. The additional structure visible very close to the centre $R \lesssim 2$~kpc is caused by the collapse of ambient gas along and around the spin axis towards the galactic plane (see Appendix~\ref{s:stab}).

To sum up, we do not find any significant differences between the vertical response of the stellar disc in a pure N-body simulation and a simulation that includes a light, cold gas disc. We find that the median and maximum vertical displacement and vertical velocity both in the stellar disc and the gaseous disc are consistent with observations during a broad time span after the impact, but that they do dampen with time. The damping of the corrugations is strongest with decreasing galactocentric radius, and it is stronger for the gas waves compared to the stellar waves.

\subsubsection{Vertical specific energy} \label{s:ener}

\begin{figure*}
\centering
\includegraphics[width=0.32\textwidth]{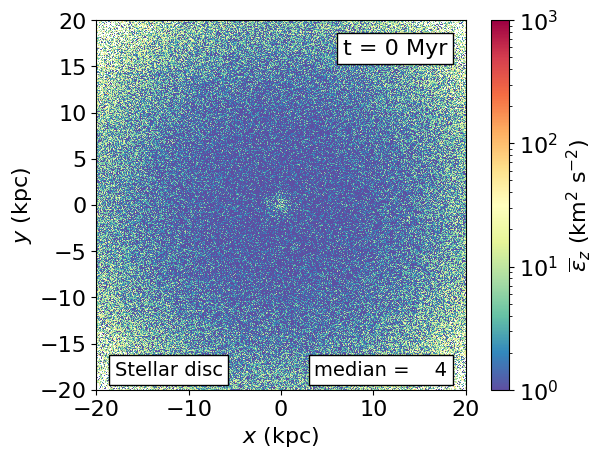}
\includegraphics[width=0.32\textwidth]{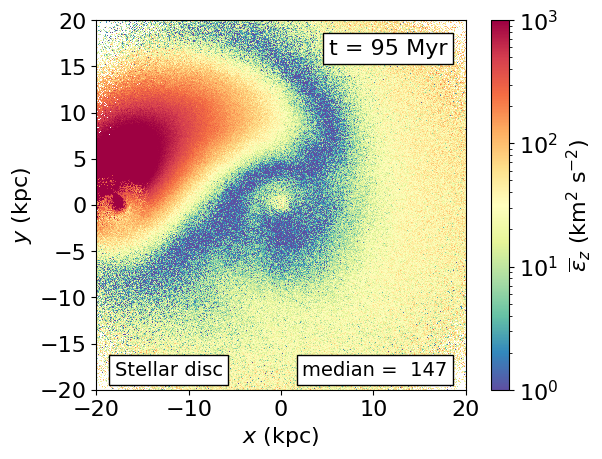}
\includegraphics[width=0.32\textwidth]{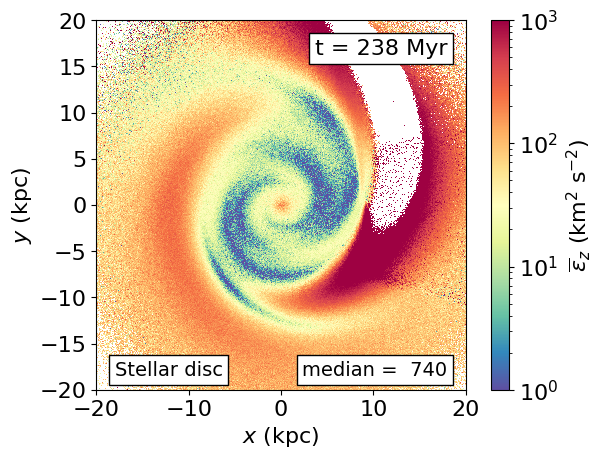}
\includegraphics[width=0.32\textwidth]{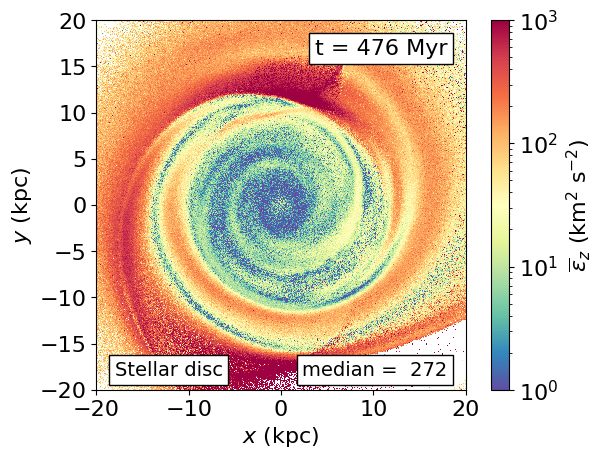}
\includegraphics[width=0.32\textwidth]{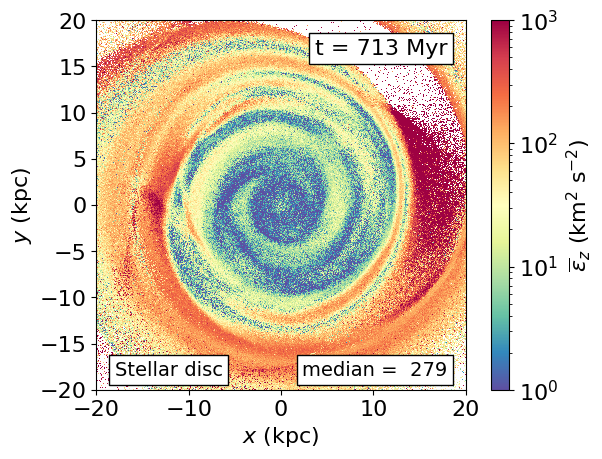}
\includegraphics[width=0.32\textwidth]{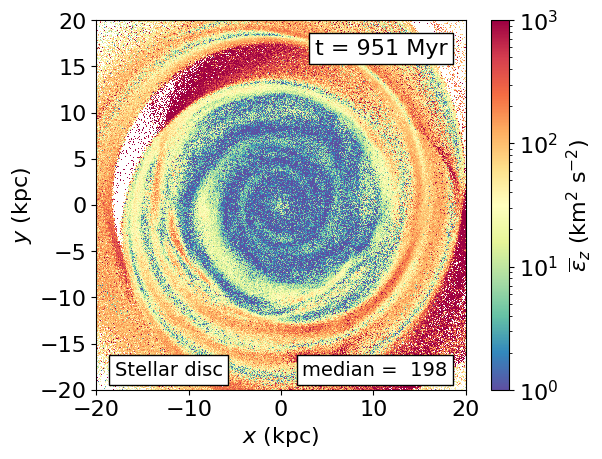}
\caption[  ]{ Vertical specific energy (Eq. \ref{e:ener_z}) across the stellar disc in the interaction hybrid simulation from roughly 100 Myr prior to the impact to roughly 850 Myr after impact. Each panel corresponds to a selected time step, indicated on its top-right corner. The colour coding indicates the value of the mean vertical energy at each location across the disc. The median value of the energy (in ${\rm km}^2 ~{\rm s}^{-2}$) across the entire disc in each snapshot is indicated on the bottom-right corner of each panel. Clearly, the vertical energy rises significantly around the impact epoch $t \approx 100$ Myr), and it is generally higher at larger radii, regardless of the epoch. Note that the impact site is clearly visible at $x \approx -18$~kpc in the central panel, top row.
}
\label{f:star_ener_z_inter}
\end{figure*}

An alternative -- and complementary -- view to the evolution of the stellar and gas corrugations in terms of their spatial and kinematic amplitudes is based on the analysis of the vertical energy involved in the process. We now pursue this complementary approach.

First, note that squared-length of the vector $w$ defined earlier (Sec.~\ref{s:pcorr}) corresponds to the total energy of a simple harmonic oscillator with frequency $\nu$ and motion constrained to the $z$-axis,
\begin{equation} \label{e:ener_z}
	\epsilon_z = \frac{ 1 }{ 2 } \left( \nu^2 \overline{z}^2  + \overline{V}^2_z \right)  \, .
\end{equation}
This expression is approximately equal to the vertical specific energy of an ensemble of stars. Based on the results discussed in Sec.~\ref{s:amp}, we assume that it can also be applied to the gas phase. We stress that when applying the above equation we are implicitly {\em considering only the bulk motions} of temporarily co-spatial ensembles of stars or patches of gas found in a small volume. We are also ignoring internal random motions or any momentary changes due to compression or expansion processes in the gas.\footnote{ Recall that the gas is kept isothermal and thus heating / cooling processes are perfectly balanced, by design. } The distinction between the energy involved in the corrugation (i.e. the bulk motions), and the actual orbital energy (of stars) is important, as will become clear later.

\begin{figure*}
\centering
\includegraphics[width=0.32\textwidth]{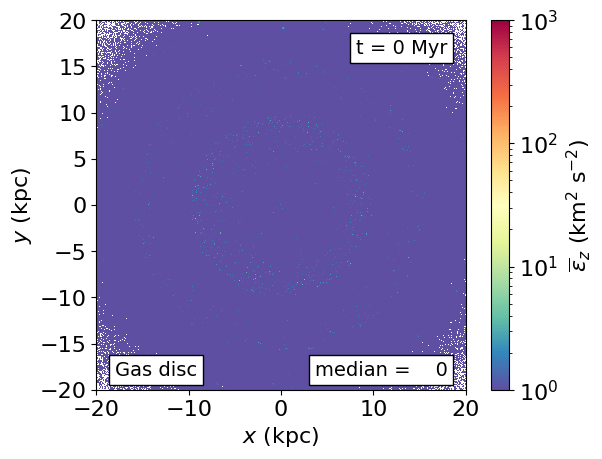}
\includegraphics[width=0.32\textwidth]{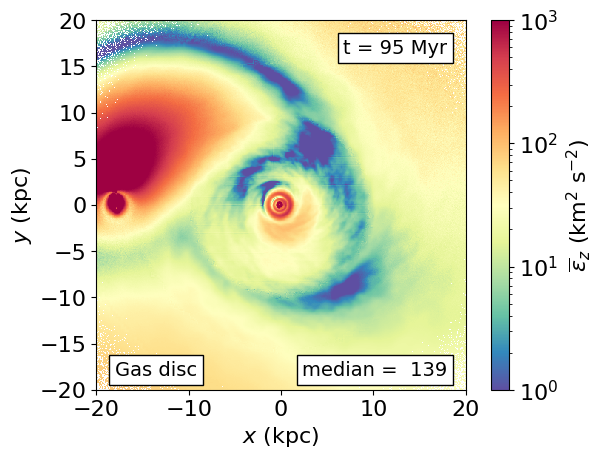}
\includegraphics[width=0.32\textwidth]{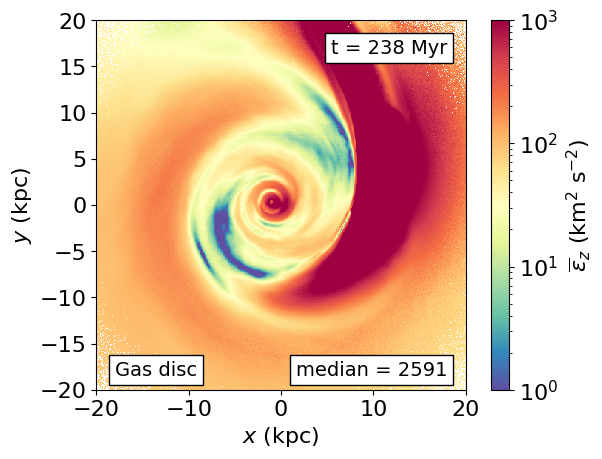}
\includegraphics[width=0.32\textwidth]{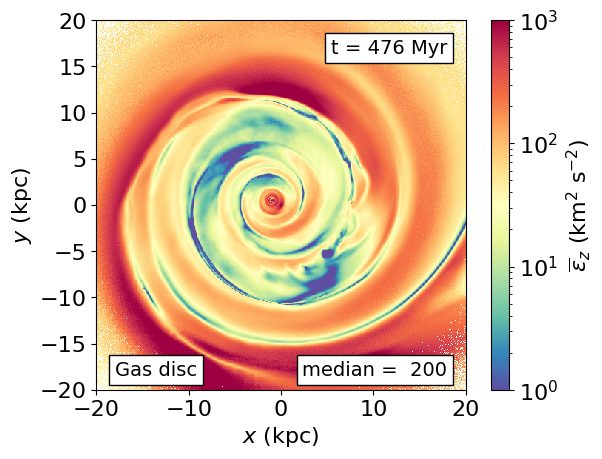}
\includegraphics[width=0.32\textwidth]{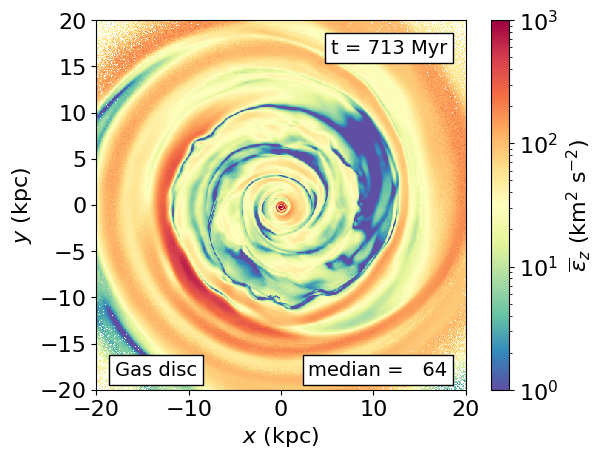}
\includegraphics[width=0.32\textwidth]{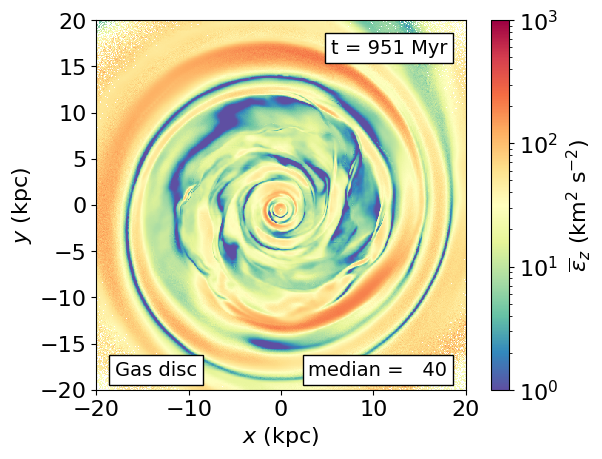}
\caption[  ]{ Same as Fig.~\ref{f:star_ener_z_inter}, but for the gas disc. Note that the impact site is clearly visible at $x \approx -18$~kpc in the central panel, top row.
}
\label{f:gas_ener_z_inter}
\end{figure*}

Using the same approach used to construct the \mbox{$\overline{z}$} and $\overline{V_z}$ maps earlier (Sec.~\ref{s:inter}), we construct separate \mbox{ $\overline{\epsilon_z}$} maps for the stars in the disc and for the gas disc. Figs.~\ref{f:star_ener_z_inter} and \ref{f:gas_ener_z_inter} display examples of such maps at selected snapshots throughout the evolution of synthetic galaxy in the interaction hybrid model. At $t = 0$ Myr (top left panel in either figure), the vertical energy is very low both across the stellar disc and the gas disc. In the absence of an external perturbation, the energy remains virtually constant in the stellar disc, and reasonably constant in the gas disc, in particular in the outer disc (cf. Fig.~\ref{f:star_ener_z_iso} and \ref{f:gas_ener_z_iso}). The behaviour of the stellar disc in the interaction N-body simulation is qualitatively similar to that of the stellar disc in the interaction hybrid simulation and its corresponding \mbox{ $\overline{\epsilon_z}$} maps are therefore omitted here.

But the interaction with a massive perturber, on the other hand, leads to a dramatic increase in the energy across both discs. The stars and the gas near the impact site at $x \approx -18$~kpc experience a significant energy transfer in the vertical direction as a result of the direct collision with the perturber. The signature in the energy increase is consistent with the onset of an $m = 1$ bending mode. The stars and the gas on the opposite side also display a significant increase in their energy, as a result of the early perturbation induced by the satellite as it its on its way towards the galactic plane. Overall, the behaviour in the increase of the vertical energy across the discs mimics the behaviour we observe in the vertical displacement and in the vertical velocity (cf. Figs.~\ref{f:z_mean} and \ref{f:vz_mean}), which is not surprising -- but consistent, given the relation between $\epsilon_z$, $z$, and $V_z$ (Eq. \ref{e:ener_z}).

\begin{figure}
\centering
\includegraphics[width=\columnwidth]{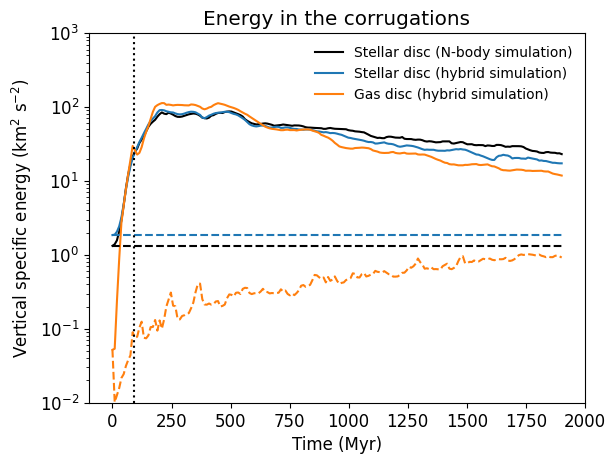}
\includegraphics[width=0.97\columnwidth]{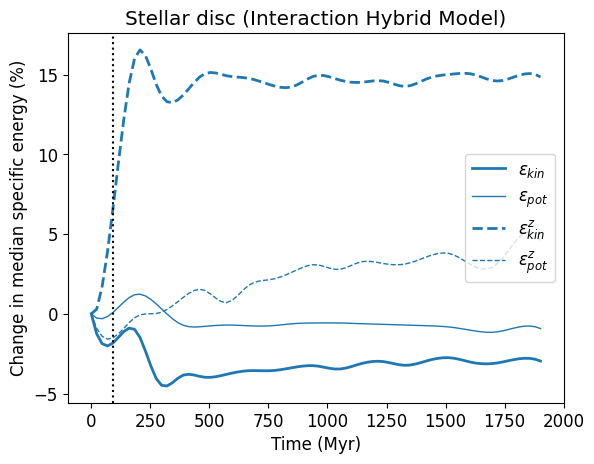}
\caption[]{
Top: Evolution of the median vertical specific energy of the corrugation in the gaseous disc (continuous orange), in the stellar disc in interaction hybrid simulation (continuous blue), and in the stellar disc in interaction N-body simulation (continuous black). The time baseline extends from roughly 100 Myr prior to the impact (flagged by the vertical dotted line) to roughly 1800 Myr after impact. For comparison, we include the results for each component in the corresponding isolated simulation (dashed curves). Note that the latter are nearly constant throughout for the stellar disc -- as expected for a stable isolated galaxy model, while it rises steadily for the gas disc, mainly as a result of the collapse of gas along the galaxy's spin axis (see Appendix~\ref{s:stab}). 
Bottom: Evolution of the change (in percent) in the median value of the orbital energy of individual disc stars in the interaction hybrid simulation, averaged over the entire disc. The total energy is split into total kinetic energy (solid thick), total potential energy (solid thin). We further show the contribution of the vertical kinetic energy (dashed thick) and vertical potential energy (dashed thin). Compare the latter to the evolution of the vertical energy of the stellar corrugation shown in the top panel. The evolution of the total orbital energy of the stars in the interaction N-body simulation are very similar, and it is therefore omitted.
}
\label{f:ener_evol}
\end{figure}

In addition to the average vertical energy {\em at each location} on the stellar and gaseous discs, we estimate the evolution of the average vertical energy by measuring the median value of the vertical energy {\em across the entire disc}, separately for the stellar disc and for the gas disc, at each time step over the full time span of the simulation. This is useful to quantify the global, long-term behaviour of the stellar and gas corrugations. The result is presented in Fig.~\ref{f:ener_evol} (top panel). The blue continuous curve indicates the evolution of the energy of the stellar disc in the interaction hybrid model, while the orange continuous curve indicates the corresponding result for the gas disc. For reference, we include the evolution of vertical energy of each disc in the isolated hybrid simulation (dashed curves). In this simulation, the vertical energy of the stellar disc remains virtually constant. The vertical energy of the gas disc is initially practically zero (consistent with the fact that the disc is initially in vertical hydrostatic equilibrium), and increases slowly with time - as a result of the collapse of gas onto the disc plane (see Appendix~\ref{s:stab}), but remains very low overall.

Comparing the blue continuous curve to the blue dashed curve, and the continuous orange curve to the orange dashed curve, we see that there is a clear excess in vertical energy in both discs as a result of the impact (flagged by the vertical dotted line) with respect to a no impact scenario, in line with \citet{ede97b}'s findings.

\begin{table}
\begin{center}
    \begin{tabular}{l|r|r}
         Simulation & Component & $\tau$ (Myr) \\
         \hline
         Interaction N-body & Stellar disc & $1350 \pm 17$\\
         Interaction hybrid & Stellar disc & $1051 \pm 14$\\
         Interaction hybrid & Gas disc & $802 \pm 15$
    \end{tabular}
\end{center}
\caption{ Corrugation damping rates (see text for details). }
\label{t:tau}
\end{table}

To asses the potential effect of the gaseous disc on the stellar disc, we include in the top panel of Fig.~\ref{f:ener_evol} the evolution of the vertical energy in the stellar disc for the interaction N-body model (black continuous curve), and the corresponding result for the isolated N-body simulation (black dashed curve).

We note that in all cases the vertical energy decreases with time, but at different rates, as suggested by the difference between the continuous curves towards the end of the simulation. By fitting a simple exponential $\propto \exp{[-t/\tau]}$ to the evolution of the vertical specific energy of each of the discs from $t = 600$ Myr onward\footnote{This choice is motivated by the fact that all three curves intersect roughly at $t = 600$ Myr. } we learn that the discs each lose vertical energy on a typical time scale $\tau \approx 1350 \pm 17$ Myr (stellar disc, interaction N-body model), $\tau \approx 1051 \pm 14$ Myr (stellar disc, interaction hybrid model), and $\tau \approx 802 \pm 15$ (gas disc). These results are summarised in Tab.~\ref{t:tau}. Thus the corrugations in gas disc disc dampen at a faster pace, compared to the stellar disc. Specifically, between $t = 600$ Myr and $t = 1800$ Myr, the energy of the corrugations in the interaction hybrid simulation have declined by roughly 68 percent (stellar disc) and roughly 78 percent (gas disc).

\begin{figure*}
\centering
\includegraphics[width=0.32\textwidth]{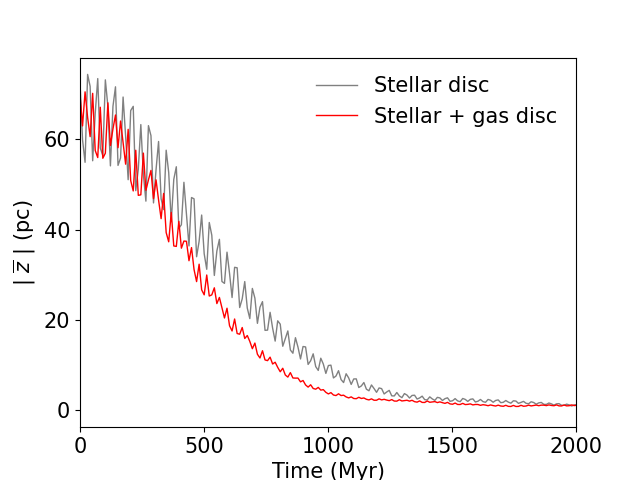}
\includegraphics[width=0.32\textwidth]{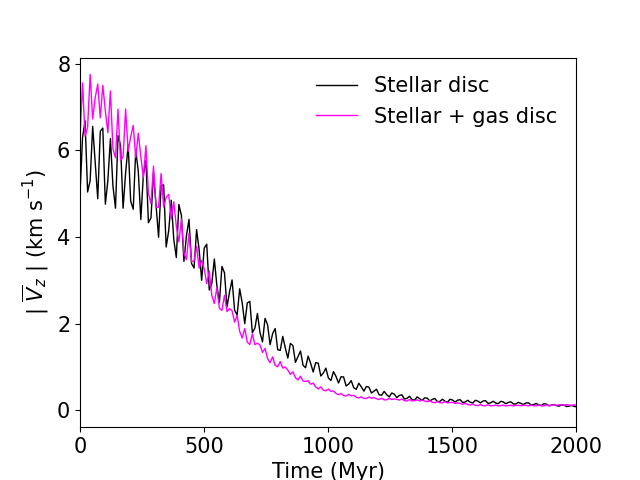}
\includegraphics[width=0.32\textwidth]{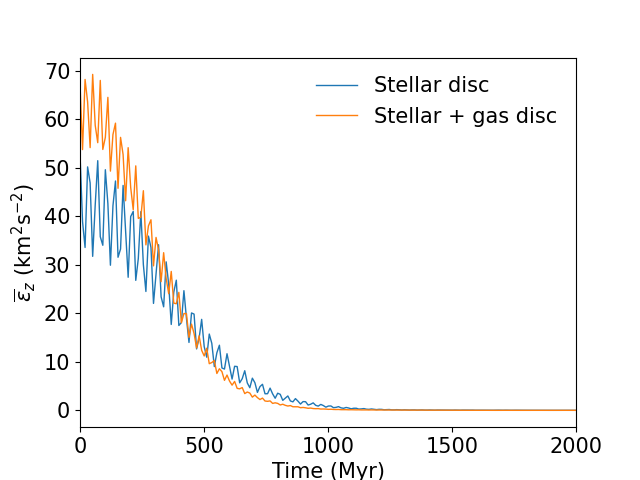}
\caption[]{ Evolution of the mean vertical amplitude (left panel), mean vertical velocity (central panel), and mean vertical energy (right panel) of an ensemble of stars undergoing phase mixing. Since the values of $\overline{z}$ and $\overline{V}_z$ oscillate with high frequency around zero, we show instead the moving average of the absolute value in each case, calculated with a box-car window of width equivalent to 40 Myr. A stronger potential close to the midplane, originating from the combined effect of a stellar and a gaseous disc, results in a stronger (faster) damping of the spatial and kinematic amplitudes, and of the vertical energy, compared to a stellar-only potential. See text for details.}
\label{f:phase_mixing}
\end{figure*}

\subsection{Wave damping} \label{s:mix}

It is not at all obvious what causes the slow damping of the corrugations observed in our simulation. We have carried out an extensive study of both the spatial and kinematic amplitudes individually (Sec. \ref{s:amp}), and the vertical energy of the undulation (Sec.~\ref{s:ener}). We are able to rule out energy being lost to the dark matter halo. In our view, there are only two possible explanations: 1) the ordered, vertical energy imparted to the stars (and in consequence to the gas) by the interaction is slowly converted into in-plane (random) kinetic energy; or 2) incomplete phase mixing.

The conversion of ordered vertical energy into random (in-plane) motion is possible if stars scatter off DM particles \citep[a numerical artefact, e.g.][see also \citealt{lac85m}]{lud21p}. It is also possible that gas clumps form during the course of the simulation that may have the same effect. Also, stars may scatter off (stellar and gaseous) spiral arms \citep{mas97b}. However, none of that happens in the simulation, or if it does, it is not relevant to the damping of the corrugations, as the following analysis suggests.

We calculate the evolution of the total energy in our simulations for all components. We find that the change in the total energy (i.e. across all components) in all simulations is on the order of less than 0.1 per cent over $\sim 2$ Gyr, i.e. energy is well conserved. This result alone is however not enough to answer our question, as it still allows for energy transfer between components, e.g. between the stellar disc and the surrounding dark matter. Thus we turn our attention to the evolution of the orbital energy of individual stars across the stellar disc.

The energy of the individual stars in the disc, split the total energy into kinetic energy, is given by
$
	\epsilon_{kin} =  \frac{1}{2} ( V_x^2 + V_y^2 + V_z^2 )
$
and potential energy
$
	\epsilon_{pot} = \Phi(x,y,z) \, .
$
The phase-space coordinates $(x, y, z)$ and $( V_x, V_y, V_z )$ are the position and velocity of a star relative to the galaxy's CoM and $\Phi$ is the total gravitational potential at a given time. Furthermore, we split each of the kinetic and potential energies into in-plane (`$ip$') and vertical (`$z$') contributions,
\begin{align} \label{e:ener}
	& \epsilon^{ip}_{kin} \equiv \frac{1}{2} ~\left( V_x^2 + V_y^2 \right) \quad ; \quad
	\epsilon^z_{kin} \equiv \frac{1}{2}~V_z^2 \\
	& \epsilon^{ip}_{pot} \equiv \Phi(x,y,0) \quad ; \quad
	\epsilon^z_{pot} \equiv  \Phi(x,y,z) - \Phi(x,y,0) \, .
\end{align}
The total energy of a star's orbit is given by the sum of these four individual contributions. It is worth noting that quantifying the in-plane and vertical contributions to the total potential by $\epsilon^{ip}_{pot}$ and $\epsilon^z_{pot}$, respectively, is a valid approach given that the overwhelming majority of stars away from the galaxy's centre (i.e. $R \gtrsim 5$ kpc) are moving along roughly circular orbits, and thus the potential is separable in $R$ and $z$ \citep[cf.][their section 3.2]{bin08a}.

The median values of the kinetic and potential energies of the stellar disc change by no more than 5 and 1 percent relative to their initial value as a result of the interaction, but then remain fairly steady for the remainder of the simulation (see Fig. ~\ref{f:ener_evol}, bottom panel, continuous curves). The same behaviour is displayed by the stellar disc in the N-body simulations and the corresponding results are therefore omitted from our discussion.

We note that {\em prior} to the impact there is a noticeable dip relative to the initial value in the kinetic and potential energies of the disc. This is likely a consequence of numerical noise which presumably leads to a readjustment of all components. Our suspicion is supported by the same behaviour being observed in the isolated hybrid simulation (cf. Fig.~\ref{f:ener_evol_iso}). The subsequent bump and dip in the kinetic energy can be understood as a consequence of the impact, and of the system's revirialisation, respectively. Indeed, during the (impulsive) interaction an energy $\Delta \epsilon_{kin}$ is added to the system's initial kinetic energy, $\epsilon_{kin,~0}$ over a relatively short timescale. At revirialisation, the kinetic energy of the system is  $\epsilon_{kin,~0} - \Delta \epsilon_{kin}$, i.e. $2  \Delta \epsilon_{kin}$ {\em lower} than the kinetic energy shortly after the impact \citep[e.g.][]{bin08a}.

We now look at the vertical energy of individual stars across the disc (see Fig. ~\ref{f:ener_evol}, bottom panel, dashed curves). The median value of the vertical kinetic (thick dashed) increases significantly by 15 percent relative to its initial value as a result of the interaction (cf. Fig.~\ref{f:ener_evol_iso}) but remains steady after that. The median value of the potential energy (thin dashed) increases slightly with time by about 5 percent relative to its initial value after roughly 2 Gyr.

The increase in the latter is modest, even more so in view of the fact that the vertical energy is orders of magnitude smaller than the  in-plane energy. Therefore, our energy analysis indicates that the average vertical energy of individual stars is reasonably well conserved during the course of the simulation after the interaction. This result is clearly at odds with the significant decrease of the vertical energy we observe in the corrugations of the stellar disc ($\sim 68$ per cent; see previous section). We conclude that the latter cannot be explained in terms of energy transfer between galaxy components, or in terms of the conversion of ordered energy into random energy, and an alternative explanation is required.

Thus we consider now our second hypothesis: that the wave damping can be explained in terms of the individual stars undergoing phase mixing. Pending a more thorough analysis, we employ here a toy model for phase mixing previously introduced in a different context \citep[][see also \citealt{can14c}]{ant18b}. In brief, we consider and ensemble of disc stars initially in dynamical equilibrium which have been vertically perturbed, both spatially and kinematically. We assume these stars behave as test particles that oscillate with respect to their equilibrium position within an anharmonic potential $\Psi$, such that their vertical frequency,
\mbox{$
	\Omega_z = \pi / 2 t_{\rm ff} \, ,
$}
 depends on the vertical amplitude $A$ of their orbit, where
$$
	t_{\rm ff}(A) = \frac{1}{\sqrt{2}} \int_0^A \left[ \Psi(A) - \Psi(z) \right]^{-1/2} dz
$$
is the free-fall time of a star from $z = A$ to $z = 0$ that initially at rest \citep[][]{can14c}. Note that $\Psi$, and thus $t_{\rm ff}$ and $\Omega_z$, depend on both $R$ and $A$. Our toy model follows closely the setup presented by \citet{ant18b}. In particular, we adopt a \citet{miy75a} disc potential, defined by $M = 10^{11}$ \Msun, and scale parameters $a = 6.5$ kpc and $b = 0.26$ kpc. We refer the reader to \citet[][their section `Methods' ]{ant18b} for further details.

Using this toy model, we simulate the evolution of an ensemble of $10^5$ stars from $t = 0$ to $t = 2$ Gyr, and calculate at each time step $\Delta t = 10$ Myr the absolute value of the mean vertical displacement $\overline{z}$, of the mean vertical velocity $\overline{V_z}$, and the value of the mean vertical energy $\overline{\epsilon}_z$ of an ensemble of stars. The results are shown in Fig.~\ref{f:phase_mixing} by the grey (left panel), black (central panel), and blue (right panel) lines, respectively. Clearly, the amplitude of both the vertical displacement and the vertical velocity, and the vertical energy decrease nearly exponentially with time $-$ this is a signature of wave damping. However, the energy of the {\em individual} stars is well conserved throughout (not shown), because there are no energy sinks or sources of any kind in this toy model. This situation is qualitatively similar to what we observe in our simulations, in particular the contrast between the damping of the corrugation on the one hand, and the conservation of the orbital energy of individual stars on the other. The decrease of $\overline{z}$ and $\overline{V_z}$ with time in our toy model can be qualitatively understood as a consequence of the distribution of stars on the $(z, V_z)$ plane evolving into a more symmetric configuration. The decrease of the vertical energy is a consequence of the latter and the relationship between energy and the phase-space variables (Eq.~\ref{e:ener_z}).

In order to test whether this simple model can explain the difference, we see between the damping of the stellar disc in the N-body simulation and the hybrid simulation (cf. Figs.~\ref{f:gas_star_amplitude} and ~\ref{f:ener_evol}, top panel, and Tab.~\ref{t:tau}), we proceed as follows. Our hypothesis is that the additional potential contributed by the gaseous disc is the reason for the stronger damping in the stellar disc (hybrid simulation). Thus we simulate the evolution of the stars using our toy model, but now increase the strength of the potential $\Psi$. To enhance the effect, we increase the potential depth by 50 percent relative to its initial value, i.e. now we adopt $M = 1.5 \times 10^{11}$ \Msun.
The evolution of  $| \overline{z} |$, $| \overline{V}_z |$, and $\overline{\epsilon}_z$ in this stronger potential is shown in Fig.~\ref{f:phase_mixing} by the red (left panel), magenta (central panel), and orange (right panel) lines, respectively. Clearly, the damping in each case is stronger compared to the weaker potential (grey, black, blue). Again, this is similar to what we observe in our simulations (cf. Figs.~\ref{f:gas_star_amplitude} and \ref{f:ener_evol}). Physically, a potential which is deeper at the midplane leads to a stronger (faster) damping as a consequence of a higher vertical frequency. Indeed, if the potential scales with $\Psi_0$ and everything else being equal, then $\Omega_z \propto  \sqrt{\Psi_0}$.

Our toy model is rudimentary for several reasons. First, the adopted potential $\Psi$ does not reflect the true potential $\Phi$ in our simulations. In particular, the `stellar+disc' potential is several times stronger\footnote{ The mass ratio of the gas disc to stellar disc in our simulation is modest, roughly 1:10. } than the actual stellar disc+gas disc potential in our simulation. More importantly, a simple 1D model neglects many important aspects of a full 3D simulation, notably the self-gravity of the disc. Yet, we believe that the model captures the most basic behaviour of an ensemble of stars in a small volume in our simulation.

Thus, while recognising that a more rigorous treatment is needed, for the time being we conclude that the damping of the stellar corrugations, as well as the result that the damping is stronger in the presence of a gas disc, {\em can be explained in terms of the stars undergoing phase mixing}. The damping of the gas corrugations, on the other hand, is likely a result of the dissipative nature of the gas, as well as of the absence of a continuous driving force (perturbation).

The stellar corrugations can thus be regarded as an {\em emergent} phenomenon, one that arises from the collective and coherent (ordered) motion of a co-spatial ensemble of stars found in a small volume, and one that is in contrast to the orbital motions of individual stars.

\section{Discussion} \label{s:discuss}

The subject of disc perturbations in galaxies has a long history. Even with the simplifying assumption of an infinitely thin disc, the phenomenology is highly non-linear because of the disc's differential rotation, inter alia \citep[e.g.][]{hun69l}. This forces us to embrace large numerical simulations to make progress, supported by an increasingly elaborate theoretical framework \citep[e.g.][]{fou15w}.

Corrugations triggered by a passing perturber have been investigated for several decades, generally confined to stellar discs, with a few noteworthy exceptions \citep{gom17a,lap18a}. The perturber triggers a complex interplay between a spiral density wave and a bending wave, both of which wrap up but at different angular rates \citepalias{bla21e}. The highest resolution simulations reveal this to be an  extraordinarily complex phenomenon with several distinct pattern speeds (timescales) at play. It is, therefore, unsurprising that adding a gaseous medium enhances the complexity further. However, this is not the case at early times, at least in our simulations, where the gas and stars are seen to evolve together. The timescale over which the stellar and gas corrugations are in phase and strong ($\sim 500-700$ Myr) is similar to the typical timescale required by the onset and evolution of phase spiral \citep[\citetalias{bla21e}; ][]{bin18a,lap19a}. At late times, the gas and stars go their separate ways.

There has been very little theoretical work on the interaction of a corrugated gas wave with a spiral density wave. In light of early observations of the Milky Way's gas layer, \citet{nel84y} looked at this scenario. First, he demonstrated the well-known 1D result of how gas overrunning a spiral density wave is compressed in the reference frame of the density wave. There is extensive evidence for mild shocks, in particular, to explain the confinement of the dust along the spiral arm.
Secondly, with the aid of a simple undulating model and assuming the tight-winding approximation, he examined how this 1D compression evolves as the bending wave and spiral density wave move in and out of phase. In their simple shock picture, the compressed gas splits into two components that come together again when the bending mode and spiral wave are in lockstep again. The suggested timescale for the splitting is of order $\sim 100$ Myr. We do not see this behaviour: this may be because the density wave `rides' the corrugation wave rather than being confined to the midplane (\citetalias{bla21e}), an outcome that was not foreseen by \citet{nel84y}.

Quite apart from the gas, we are unable to find earlier research on the interaction of the {\it stellar} corrugation and the stellar density wave. The first clues to this complex interaction have come from our study of the phase spiral evolution across the disc \citepalias{bla21e}.

In our simulations, we observe a damping of the stellar and gaseous corrugation, albeit with different timescales. A toy model is used to demonstrate that the damping of the stellar corrugations can be understood in terms of an ensemble of stars undergoing phase mixing, which results in their distribution on vertical phase space becoming more symmetric around the origin with time \citep[cf.][]{kha21a}. A stronger disc potential, as originating from the combined effect of a stellar and a gaseous disc, results in a shorter damping timescale. The damping of the gaseous corrugations is, we believe, a consequence of the dissipative nature of the gas \citep[cf.][]{mos10a}, and the absence of a continuous driving force.

At the present time, we cannot easily relate our simulations to the observations. There has been no systematic survey of the 
local topography of the gas corrugation across the Galactic disc, or even its orientation with respect to the stellar corrugation or spiral arms, although there are individual efforts. \citet{alv20a} has traced local gas clouds ($\sim 3\times10^6$ \Msun) in a spatially and kinematically coherent, undulating structure ($\sim 2$~kpc average period, 160 pc maximum amplitude) with an aspect ratio of 1:20, roughly 2.5~kpc thick. The `Radcliffe wave' appears to be tangential to the radius vector to the Galactic centre but this only occurs at late times in our simulations when both the bending mode and density wave are tightly wound \cite[see also the discussion in][]{thu22p}. And there is a recent claim of the discovery of a stellar counterpart to the Radcliffe wave \citep[the so-called `Cepheus spur';][]{pan21c}.

The fact that stellar and gas disc corrugations coexist in the Milky Way may serve to rule out some of the early ideas around magnetic fields and Parker instabilities \citep[cf.][]{han92f}. The idea of magnetic fields was posed after \citet{wei91a} found that high order corrugations were not excited by a passing satellite. In fact, as we show, this appears to be true; the higher order corrugations arise from the wrapping up of the low order warp \citepalias{bla21e}. Local corrugations triggered by HVCs \citep{ten87n} or dark matter subhaloes with gas \citep{nic11b} are not able to account for the large-scale extent of what is observed in stars and gas.

We believe that the time is right for an extensive survey using wide-field integral field spectrographs, in a similar fashion to how other studies have been carried out for well known disc phenomena, e.g. TIMER survey of galaxies with bars and outer rings \citep{gad18y}. It is now possible to obtain detailed observations for dozens of objects with a monolithic instrument like MUSE on the VLT focussed on the ionised gas or stars, or using the ALMA array to map a large sample of molecular gas discs \citep{ler21n}. Given the weight of evidence (Sec.~\ref{s:intro}), we suspect that the kinematic approach is likely to be the most productive over the broadest range in disc inclination angle.
Very little is known about the context of these disc corrugations, although most appear to have massive companions. Given that warps are commonplace \citep{bin92a}, and these appear to wrap up to produce corrugations \citepalias{bla21e}, it is likely that many more will be found in a systematic survey.

There is a strong case for carrying out new simulations that include feedback processes like star formation and more realistic gas phases, in particular, would be useful. Accommodating a warm gas disc is desirable because we suspect the clearest evidence of disc corrugations will come from emission-line signatures in star formation regions.
If the gas is predicted to be highly clumped, its equation of state could be rather different from our simple assumption, and its dynamical evolution could be quite different.
Quite apart from the stellar and gas corrugations, can it be reasonably demonstrated that the newly observed properties of the spiral density wave \citep[e.g.][]{eil20g} are consistent with triggering and wind-up from the impulsive action of a perturber? How much of what we observe in the disc today arises from external action rather than internal processes? These are questions that we hope to address in later papers.

\section*{Acknowledgments}
We thank the referee for a very constructive report that helped improve our paper.
We are indebted to Eugene Vasiliev for his continuing assistance with \agama. TTG acknowledges partial financial support from the Australian Research Council (ARC) through an Australian Laureate Fellowship awarded to JBH. 
We acknowledge the use of the National Computational Infrastructure (NCI) which is supported by the Australian Government, and accessed through the Sydney Informatics Hub (SIH) HPC Allocation Scheme 2022  (PI: TTG; CI: JBH).
We made use of Pynbody\footnote{\url{https://github.com/pynbody/pynbody}} and Matplotlib \citep{hun07a} -- both {\sc Python}\footnote{\url{http://www.python.org} }-based softwares -- in our analysis.
This research has made use of NASA's Astrophysics Data System (ADS) Bibliographic Services\footnote{\url{http://adsabs.harvard.edu} }.

\section*{Data Availability}

The data underlying this article will be shared on reasonable request to the corresponding author.

\bibliographystyle{mnras} 
\input{mw_sgr_corrugations.bbl}

\appendix

\section{Stability of the initial conditions} \label{s:stab}

\begin{figure*}
\centering
\includegraphics[width=0.47\textwidth]{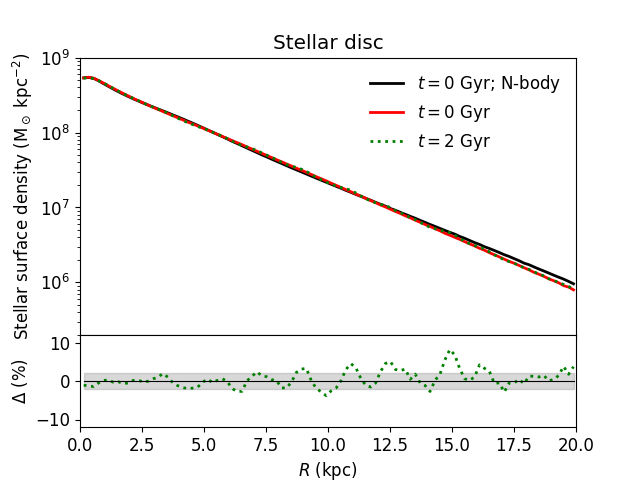}
\includegraphics[width=0.47\textwidth]{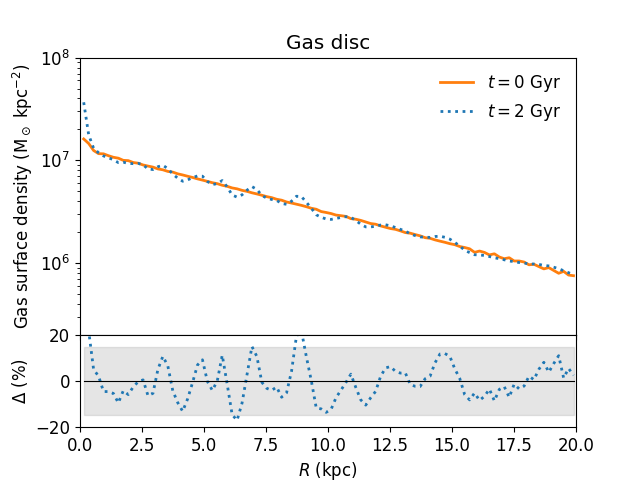}
\caption[  ]{  Stability of the initial conditions: Surface density profiles. Top. Surface density profile of the stellar disc in the isolated hybrid model at $t = 0$ Gyr (red) and $t \approx 2$ Gyr (green). For comparison, we include the surface density profile of the stellar disc in the N-body model at $t = 0$ Gyr (black curve; see \citetalias{bla21e} for further details). The bottom sub-panel shows the relative difference in percent between the model at $t = 0$ Gyr and $t \approx 2$ Gyr (green). The shaded area indicates the root-mean-square (rms) deviation for the latter case which amounts to $\pm 2$ percent. Bottom. Surface density profile of the {\em gas} disc in the isolated hybrid model at $t = 0$ Gyr (orange) and $t \approx 2$ Gyr (blue). The bottom sub-panel indicates the relative difference between the orange and the blue curves, with an rms deviation of roughly $\pm 15$ percent. See also Fig.~\ref{f:surfdens_iso}.
}
\label{f:surfdens_prof}
\end{figure*}

In this section we take a look at the stability of the initial conditions. We focus on on the structural and kinematic properties of the stellar and gas disc in the isolated hybrid model that are relevant to our study. The stability of the isolated N-body model was discussed at length in \citetalias{bla21e}.

\subsection{Structural stability}

Fig.~\ref{f:surfdens_prof} compares the surface density profile of the stellar disc (left panel) and the gas disc (right panel) at $t = 0$ and $t = 2$ Gyr in the isolated hybrid model. The top sub-panel in each of the left and right panels shows the density profile $\Sigma(R,t)$, while the bottom sub-panel shows the relative difference $\Delta \equiv [\Sigma(R,t=2) - \Sigma(R,t=0)] / \Sigma(R,t=0)$ in percent. The root-mean-squared value of the latter is $\pm 2$ percent in the case of the stellar disc, and $\pm 15$ in the case of the gas disc, across all radii between 0 and 20 kpc.

We include in the left panel of Fig.~\ref{f:surfdens_prof} the surface density profile of the stellar disc in the isolated N-body model. Our purpose is to show that the stellar discs in the hybrid and the N-body models are indeed very similar, which in turn justifies comparing their response to a crossing satellite.

We note a pile-up of gas at the centre of the disc after roughly 2 Gyr of evolution in isolation (see also Fig.~\ref{f:vz_mean_iso}, bottom middle panel, and Fig.~\ref{f:gas_ener_z_iso}, bottom row). This is commonly seen in these type of isolated, isothermal configurations \citep[e.g.][]{deg19a}, and is due to the infall of ambient gas mainly along the galaxy's spin vector. This in turn is a consequence of the ambient gas not being in strict (hydrostatic) equilibrium from the outset.

\begin{figure}
\centering
\includegraphics[width=0.47\textwidth]{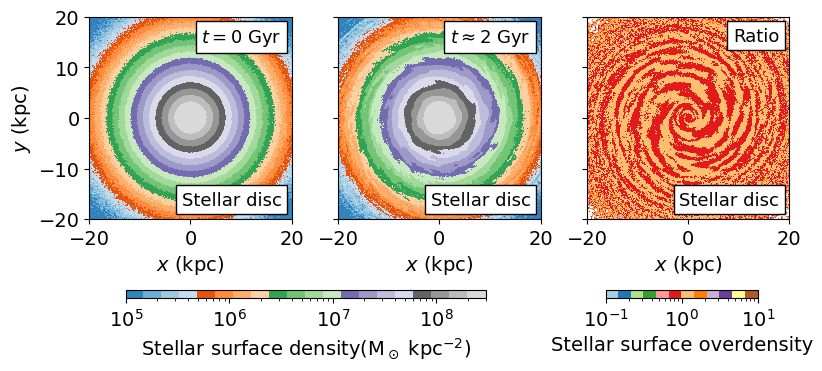}
\includegraphics[width=0.47\textwidth]{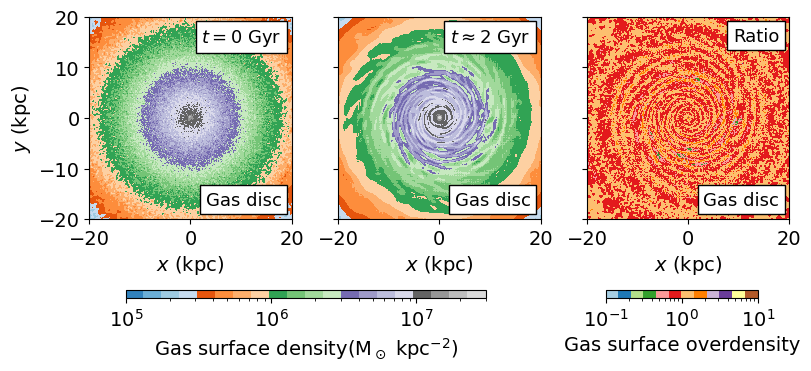}
\caption[  ]{ Stability of the initial conditions in the interaction hybrid simulation. Surface density maps of the stars in the disc (top) and of the gas (bottom) at $t = 0$ Gyr (left) and at $t \approx 2$ Gyr (middle). Note the difference in colour scale between the left and middle panels in top row and the corresponding panels in the bottom row. The right panel displays the ratio of the middle-to-left map in each case, on the same colour scale. In either case, the variations in the in the final state with respect to the initial state are at the $\sim 10$ percent level. See also Fig.~\ref{f:surfdens_prof}.
}
\label{f:surfdens_iso}
\end{figure}

The surface density profile is calculated as the azimuthal average of the surface density of the disc at each radius, and this may wash away the presence of small, non-axisymmetric substructure that develop from instabilities (noise). Therefore, a more appropriate indicator of a disc's structural stability is its surface density map, $\Sigma(x,y,t)$. Fig.~\ref{f:surfdens_iso} displays the surface density of the stellar disc (top row) and the gas disc (bottom row) in the isolated hybrid model at $t = 0$ (left column) and $t = 2$ Gyr (middle column). The right panel on each row displays the (overdensity) ratio $\Sigma(x,y,t=2)/\Sigma(x,y,t=0)$.

Both disc develop small substructure by $t = 2$ Gyr, mostly in the form of weak, flocculent spiral arms, but the overdensities across the disc are in each case of order 10 percent, which we consider acceptable for our purpose.

Fig.~\ref{f:z_mean_iso} compares the mean vertical displacement $\langle z(x,y,t) \rangle$ of the stars (top) and gas (bottom) in the disc at $t = 0$ (left column) and $t = 2$ Gyr (middle column) . The right panel on each row displays the absolute value of the difference $\Delta z \equiv \langle z(x,y,t=2) \rangle - \langle z(x,y,t=0) \rangle$. The latter reveals that changes in the initial vertical displacement of the stars due to noise are no larger than 0.05 kpc across the disc, and not larger than 0.10 kpc in the gas disc. These are significantly smaller than the typical spatial amplitude of the stellar and gas corrugations as discussed earlier.

\subsection{Kinematic stability}

Fig.~\ref{f:vz_mean_iso} displays the mean vertical velocity $\langle V_z(x,y,t) \rangle$ of the stellar and the gas discs. There it is shown that the changes in the mean $V_z$ of the stars (right top panel) and and the gas (right bottom panel) in the disc are on the order of 2 \kms\ or less, which is far smaller than the typical kinematic amplitude of the corrugations induced by a crossing satellite.

\begin{figure}
\centering
\includegraphics[width=0.47\textwidth]{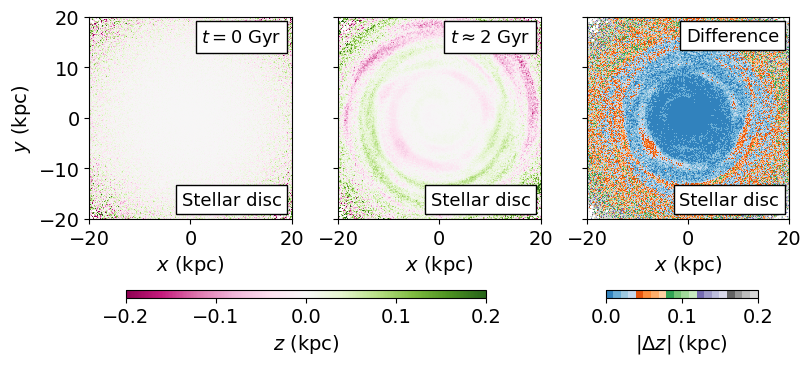}
\includegraphics[width=0.47\textwidth]{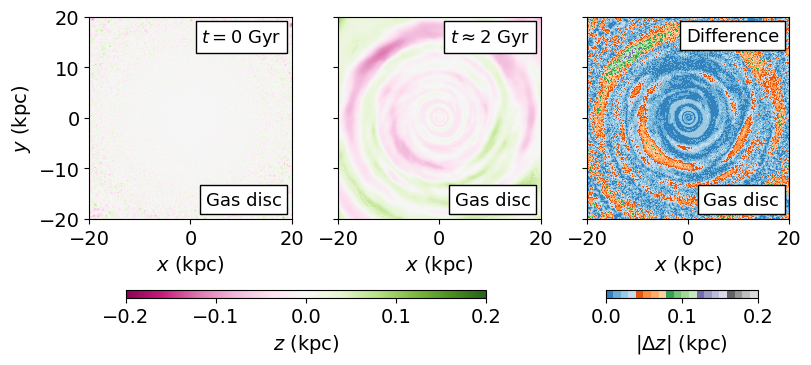}
\caption[  ]{ Stability of the initial conditions in the interaction hybrid simulation. Mean $z$ maps of the stars in the disc (top) and the gas (bottom)  at $t = 0$ Gyr (left) and at $t \approx 2$ Gyr (middle). The right panel displays the absolute difference between the middle and the left maps. Clearly, changes in the initial vertical displacement of the stars due to noise are no larger than 0.05 kpc across the disc, and not larger than 0.10 kpc in the gas disc.}
\label{f:z_mean_iso}
\end{figure}

\begin{figure}
\centering
\includegraphics[width=0.47\textwidth]{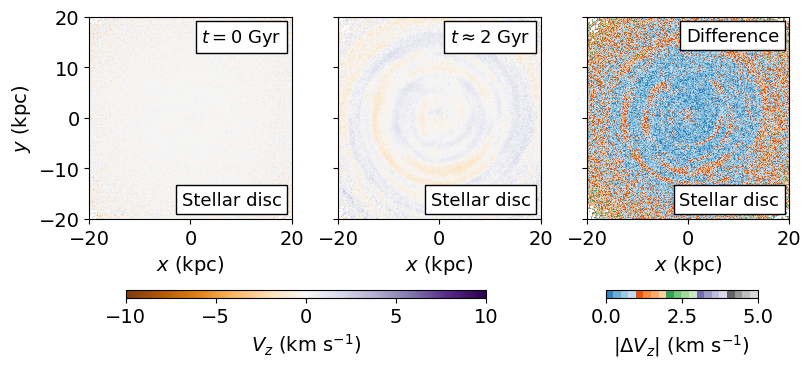}
\includegraphics[width=0.47\textwidth]{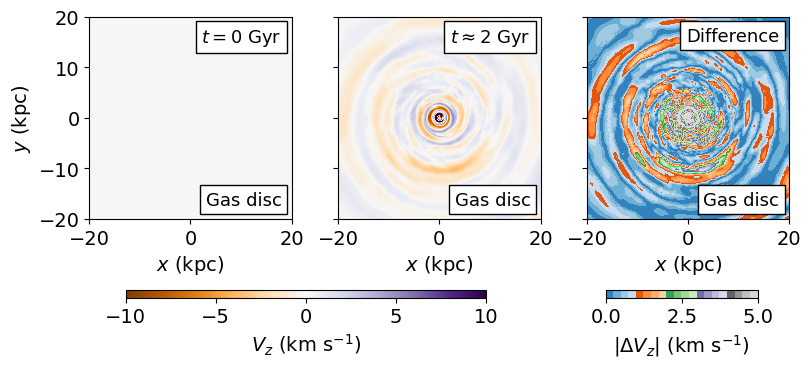}
\caption[  ]{ Stability of the initial conditions in the interaction hybrid simulation. Mean $v_z$ maps of the stars in the disc (top) and the gas (bottom)  at $t = 0$ Gyr (left) and at $t \approx 2$ Gyr (middle). The right panel displays the absolute difference between the middle and the left maps. The high amplitude around the centre at $t \approx 2$ Gyr is due to the infall of ambient, cooling gas along the axis of symmetry, which is common in simulation without heat sources or a vertically stratified atmosphere. }
\label{f:vz_mean_iso}
\end{figure}

\begin{figure*}
\centering
\includegraphics[width=0.32\textwidth]{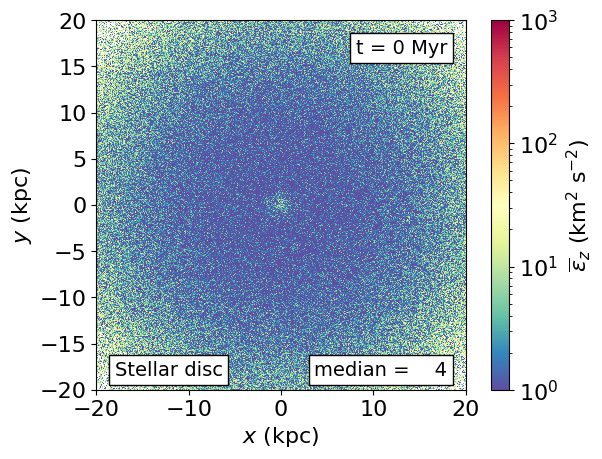}
\includegraphics[width=0.32\textwidth]{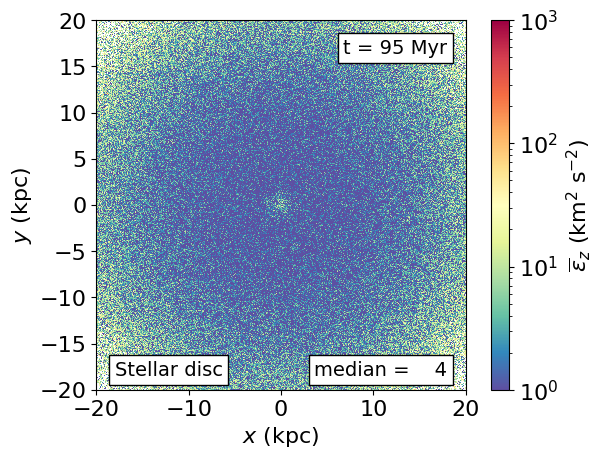}
\includegraphics[width=0.32\textwidth]{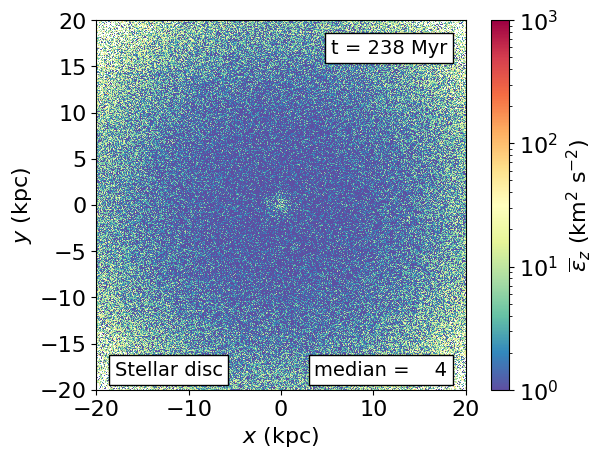}\\
\includegraphics[width=0.32\textwidth]{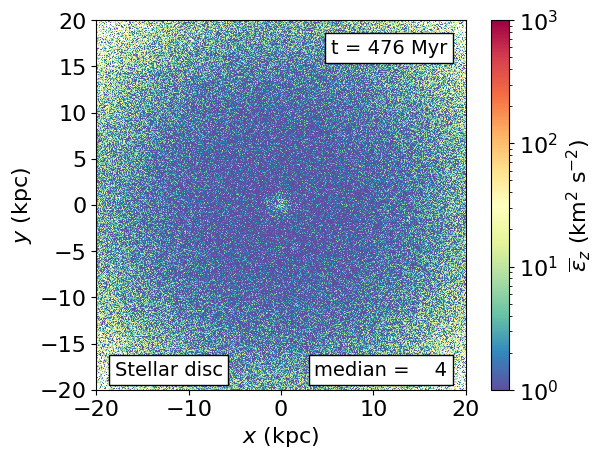}
\includegraphics[width=0.32\textwidth]{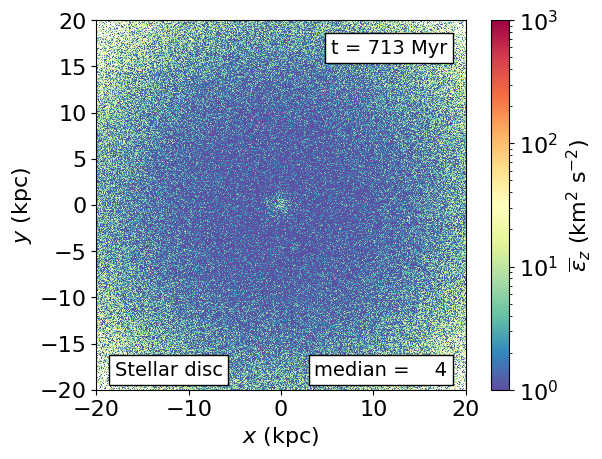}
\includegraphics[width=0.32\textwidth]{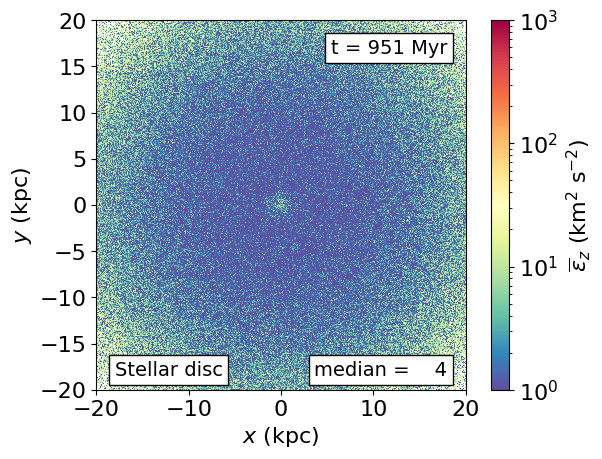}
\caption[  ]{ Mean vertical energy across the stellar disc in the isolated hybrid simulation over 1 Gyr of evolution. Each panel corresponds to a selected time step, indicated on its top-right corner. The colour coding indicates the value of the mean vertical energy at each location on the disc. The median value of the energy across the entire disc in each snapshot is indicated on the bottom-right corner of each panel. The vertical energy does not change appreciably during the time span shown, indicating that the stellar disc is in near-perfect equilibrium from the outset.}
\label{f:star_ener_z_iso}
\end{figure*}

\begin{figure*}
\centering
\includegraphics[width=0.32\textwidth]{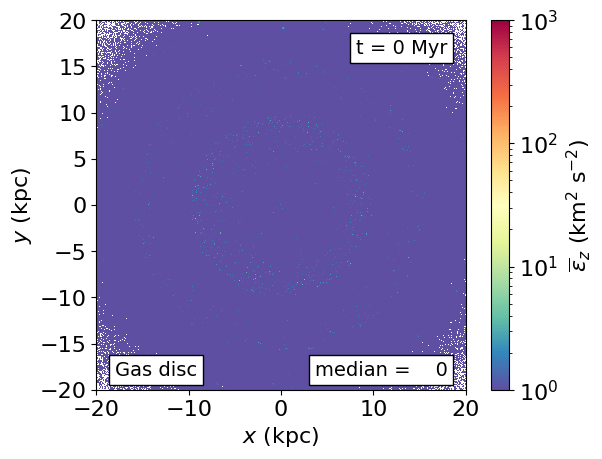}
\includegraphics[width=0.32\textwidth]{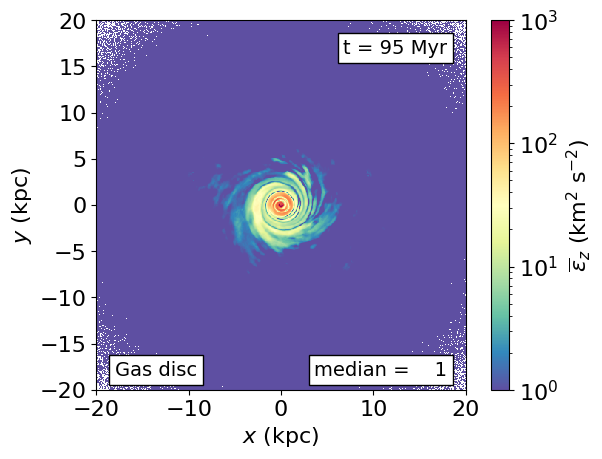}
\includegraphics[width=0.32\textwidth]{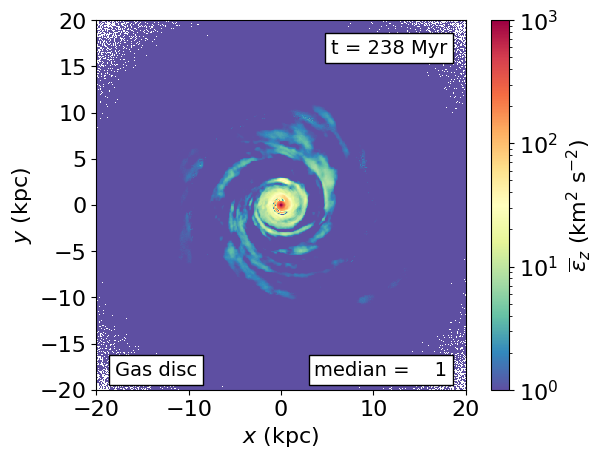}\\
\includegraphics[width=0.32\textwidth]{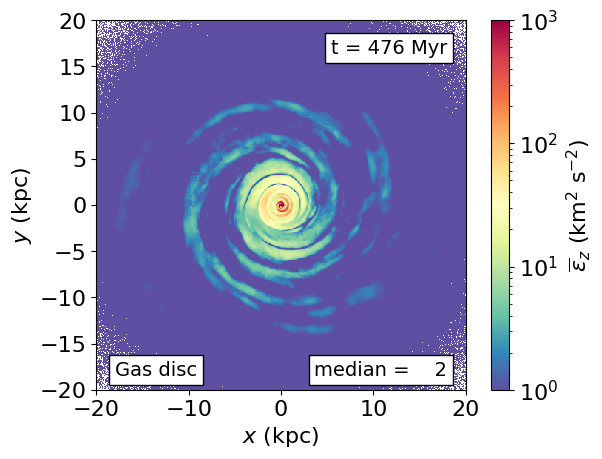}
\includegraphics[width=0.32\textwidth]{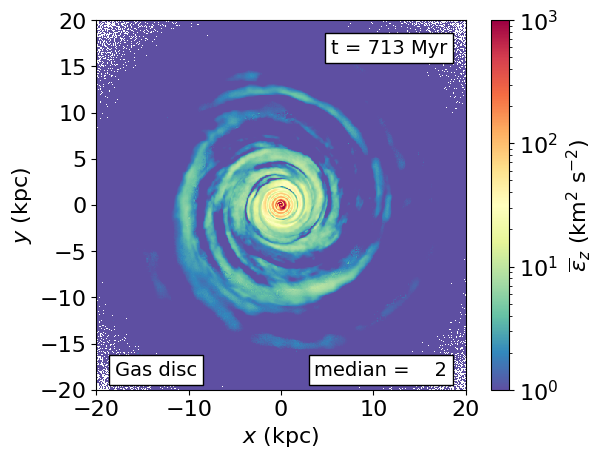}
\includegraphics[width=0.32\textwidth]{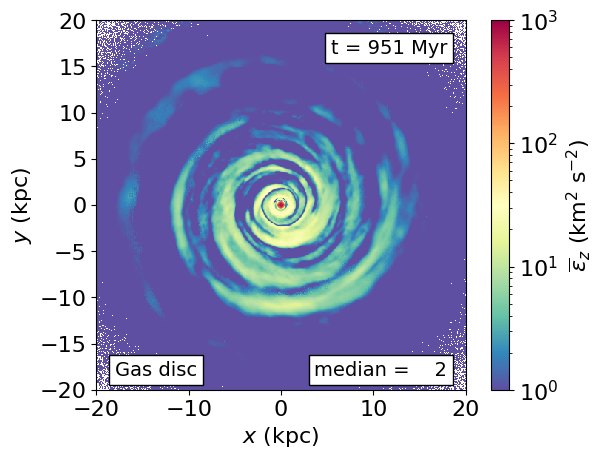}
\caption[  ]{ Same as Fig.~\ref{f:star_ener_z_iso}, but for the gas disc, and adopting a larger energy range. Although the median value of the energy across the disc remains virtually unchanged, there is some localised change in vertical energy across the disc, in particular at the centre. This is due to the collapse of the ambient gas onto the disc plane. }
\label{f:gas_ener_z_iso}
\end{figure*}

Although not strictly a kinematic property, we now look at the vertical energy involved in the development of corrugations triggered by noise. Figs.~\ref{f:star_ener_z_iso} and \ref{f:gas_ener_z_iso} show the mean vertical energy $\overline{\epsilon_z}$ of the stellar disc and the gas disc respectively. Each panel in each of these figures corresponds to a different snapshot in the simulation, spanning a range in time from $t = 0$ Gyr to roughly 1 Gyr. We find that the mean energy across the stellar disc is low, implying that the disc does not develop any significant corrugations as a result of noise. This is consistent with the results discussed earlier around Figs.~\ref{f:z_mean_iso} and \ref{f:vz_mean_iso}. In contrast, there is a noticeable increase in the vertical energy of the gas disc, particularly around the centre. However, this is not caused by the onset of powerful corrugations triggered by noise, but rather by the collapse of ambient gas onto the disc. This is supported by our finding that neither the spatial amplitude nor the kinematic amplitudes across the gas disc in isolation change significantly (cf. Figs.~\ref{f:z_mean_iso} and \ref{f:vz_mean_iso}, bottom rows). In addition, the energies involved in this case are {\em at least an order of magnitude smaller} compared to the typical energies involved in the corrugations triggered by a crossing satellite.

\begin{figure}
\centering
\includegraphics[width=\columnwidth]{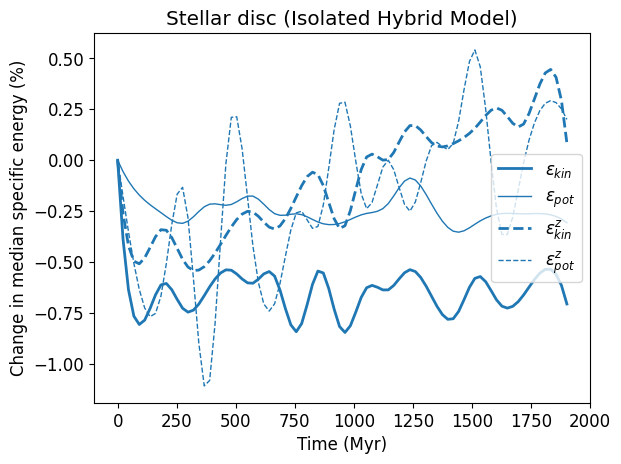}
\caption[]{  Similar to the bottom panel of Fig.~\ref{f:ener_evol}, but for the isolated hybrid simulation. }
\label{f:ener_evol_iso}
\end{figure}

Finally, we consider the energy of the stars in the disc. Fig.~\ref{f:ener_evol_iso} displays the change in the median energy split into kinetic (thick continuous curve), potential (thin continuous curve). The change of these contributions relative to their initial values is less than one percent over a time span of roughly 2 Gyr. The same is true for the contributions to the vertical energy alone, split into kinetic (thick dashed curve) and potential (thin dashed curve). These results imply that the energy of individual stars is well conserved in our simulation.

Thus we conclude that the initial conditions of our hybrid simulations are stable, and well suited for the main purpose of our paper: to investigate in detail the joint vertical response of a stellar and a gas disc to the impulse generated by a one-time crossing satellite.


\bsp	
\label{lastpage}
\end{document}